\documentclass[11 pt,draftcls]{IEEEtran} \onecolumn
\usepackage{cite}
\usepackage{algorithm,algorithmicx}
\usepackage{algpseudocode}
\usepackage[small,bf]{caption}
\usepackage[cmex10]{amsmath}
\usepackage{amssymb,latexsym,epsfig,subfigure,epic,amscd,mathrsfs,euscript,eufrak,bm}
\usepackage{color}
\usepackage{booktabs}% http://ctan.org/pkg/booktabs
\usepackage{array}% http://ctan.org/pkg/array

\usepackage{amsfonts}
\usepackage{amsopn}
\usepackage{graphicx}
\usepackage{url} 
\usepackage{empheq}
\usepackage{epstopdf}
\usepackage{float}
\usepackage{framed}

\newcommand{\argmin}{\operatornamewithlimits{arg\,min}}

\DeclareMathOperator{\diag}{diag}
\DeclareMathOperator*{\minimize}{\text{minimize}}

\DeclareMathOperator*{\st}{\text{subject to}}
\DeclareMathAlphabet\mathbfcal{OMS}{cmsy}{b}{n}
\newcommand{\Def}[0]{\mathrel{\mathop:}=}

\begin{document}
 
%\title{Memristor-Based Optimization and Learning Framework for Emerging AI Applications}
\title{A Memristor-Based Optimization Framework for AI Applications}

\author{Sijia~Liu,~\IEEEmembership{Member,~IEEE,}
 Yanzhi~Wang,~\IEEEmembership{Member,~IEEE}
       Makan~Fardad,~\IEEEmembership{Member,~IEEE,}
        and~Pramod~K.~Varshney,~\IEEEmembership{Fellow,~IEEE}% <-this % stops a space
%\thanks{Copyright (c) 2015 IEEE. Personal use of this material is permitted. However, permission to use this material for any other purposes must be obtained from the IEEE by sending a request to pubs-permissions@ieee.org.}   
\thanks{S. Liu is with the Department of Electrical Engineering and Computer Science, University of Michigan, Ann Arbor, MI, 48019 USA. E-mail: lsjxjtu@umich.edu. 
 Y. Wang, M. Fardad and P. K. Varshney are with   the Department of Electrical Engineering and Computer Science, Syracuse University, 
 E-mail: \{makan,ywang393,varshney\}@syr.edu.}
%\thanks{This work was partially supported by grants from the US Army Research Office, grant numbers W911NF-15-1-0479 and W911NF-15-1-0241. Portions of this work were presented at the $42$nd IEEE International Conference on Acoustics, Speech and Signal Processing, New Orleans, USA, in  2017.}
%\thanks{Submitted version; please do not distribute it.}
}

\maketitle

\begin{abstract}
 Memristors  have recently received significant attention as     
ubiquitous device-level
components for building a novel generation of   computing systems. 
These devices have many promising features, such as non-volatility, low power consumption, high density, and excellent scalability. 
%Their unique property to record the historical profile of the excitations on the device suggests applications in constructing super-dense non-volatile memories. However, memristors are much more than memory devices. 
The ability to control and modify biasing voltages at the two terminals of memristors make them promising candidates to perform matrix-vector multiplications and solve systems of linear equations.   In this article, we discuss how networks of memristors arranged in crossbar arrays can be used for efficiently solving  optimization  %(e.g., linear program, second-order cone program and quadratically constrained quadratic program) 
and machine learning problems. 
%(e.g., sparse coding, principal component analysis, and neural network applications). 
We introduce a new memristor-based optimization framework
%algorithm-hardware co-optimization framework 
that combines the computational merit of memristor crossbars 
%with advantages of an operator splitting method, alternating direction method of multipliers, for solving problems arising in statistics, signal processing and machine learning.  
with the advantages of   an operator splitting method, alternating direction method of multipliers (ADMM). Here, ADMM helps in splitting a complex optimization problem into subproblems that involve 
the solution of   systems of linear equations.
%such as alternating direction method of multipliers and power iteration method, for solving problems arising in statistics, signal processing and machine learning.  
The capability of this framework is shown by applying it to  linear programming, quadratic programming, and sparse optimization.  In addition to ADMM, implementation  of a customized power iteration (PI) method  for eigenvalue/eigenvector computation using memristor crossbars is discussed.  The   memristor-based PI method can further be applied to principal component analysis (PCA).
The use of memristor crossbars yields a significant speed-up in computation, and thus, we believe, has the potential to \textcolor{black}{advance} optimization and machine learning research in artificial intelligence (AI). 

%
%The answer is that memristors help us solve systems of linear equations, and ADMM helps us break up optimization problems into parts that involve solving systems of linear equations.
\end{abstract}


\IEEEpeerreviewmaketitle

\section{Introduction}
\label{sec:intro}

Memristors,   nano-scale devices conceived   by Leon Chua   in 1971, have been physically realized  by  
 scientists from Hewlett-Packard 
\cite{chua1971memristor,strsni08}.
%%% advantages of memristors
In contrast with the traditional CMOS technology, memristors can be used as %super-dense 
non-volatile memories for
building brain-like learning machines with %nano-scale 
memristive synapses \cite{kozma2012memristors}.
They  offer the ability to construct a dense, continuously programmable, and reasonably accurate cross-point
array architecture, %(known as memristor crossbar array), 
which can be used for 
  data-intensive applications \cite{hamdioui2017memristor}. 
For example, a memristor crossbar array  exhibits  a unique type of parallelism that can be utilized to perform matrix-vector multiplication and solve systems of linear equations in an astonishing $O(1)$ time complexity  \cite{huli12cf,richter2015memristive,ren2017algorithm,cai2016low}.
Discovery and physical realization of memristors 
has inspired the development of  efficient approaches to implement 
\textcolor{black}{neuromorphic computing systems 
that can mimic neuro-biological architectures and perform high-performance computing for deep neural networks and optimization algorithms} \cite{schuman2017survey}.  %\cite{schuman2017survey,huli12cf,richter2015memristive,ren2017algorithm,cai2016low,liu2015reno}. 

% A memristor crossbar array  exhibits  a unique type of parallelism that can be utilized to perform matrix-vector multiplication and solve systems of linear equations in an astonishing $O(1)$ time complexity  \cite{huli12cf,richter2015memristive,ren2017algorithm,cai2016low}.

The similarity between the programmable resistance state of memristors and the variable synaptic strengths
of biological synapses  facilitates the circuit realization of neural network models \cite{liu2015reno}.
Nowadays, an  artificial neural network   has become an extremely  popular  machine learning tool with a wide spectrum of applications, ranging from  prediction/classification,  computer vision, natural language processing, image processing, 
to signal processing \cite{lecun2015deep}. Encouraged by  its success,  
many researchers have attempted to design memristor-based computing
systems to  accelerate neural network training 
\cite{hasan15training,hasan2016reconfigurable,hasan2017chip,gokmen2016acceleration,agarwal2016resistive,li2014training,soudry2015memristor,yakopcic2017flexible,liu2016harmonica,wang2017group,ni2017energy}.
In \cite{hasan15training,hasan2016reconfigurable},
 memristor crossbars were used to form an on-chip training circuit for an autoencoder, an artificial neural network with one hidden layer. 
 Training a multi-layer neural network requires the implementation of a back-propagation algorithm  \cite{deng2014deep} for synaptic weight update.  
 Such an implementation using memristor crossbars
 was discussed 
 in \cite{hasan2017chip,gokmen2016acceleration,agarwal2016resistive,li2014training,soudry2015memristor}. 
In \cite{yakopcic2017flexible,liu2016harmonica}, a memristor-based   neural network  was proposed by using an
off-chip
training approach where synaptic weights are
pre-trained in software. This approach avoided the complexity of mapping the back-propagation algorithm onto memristors   but did not fully utilize the computational advantages of memristors.
In \cite{wang2017group,ni2017energy},  research efforts were made to
overcome    hardware restrictions, such as scalability  and   routing congestion, to design  memristor-based large neural networks.

In addition to artificial neural networks, 
memristor-based computing systems   have also been proposed and analyzed
for    sparse coding,   dictionary learning, and compressive sensing \cite{sheridan2017sparse,yu2015chip,seo2015chip,chen2015mitigating,chekad15,kadxu14,liu2017ultra}. These applications    share  a similar sparse learning framework, where   a    sparse solution is sought to minimize a certain cost function. %occurred in, e.g., model selection,  pattern recognition, and information retrieval. 
 In \cite{sheridan2017sparse},  a sparse coding algorithm  was mapped to memristor crossbars.  In \cite{yu2015chip,seo2015chip,chen2015mitigating,chekad15,kadxu14}, memristors were used to 
achieve on-chip acceleration of  dictionary learning algorithms. However, the    algorithms required the memristor network   to be programmed  multiple times  due to the gradient update step which resulted in   computation errors caused by device variations. In \cite{chen2015mitigating},  redundant memristors were   employed   to suppress these device variations.
%Different from gradient-type algorithms used in \cite{yu2015chip,seo2015chip,chen2015mitigating,chekad15,kadxu14}, the work \cite{liu2017ultra} applied ADMM to design  a  memristor-based compressive sensing system.
%It was discussed in   \cite{liu2017ultra} that 
%  ADMM programs the  memristors only once and thus the impact of    hardware
%variations is minimized.
Besides sparse learning, memristor crossbars have also been considered for implementing and training  a   probabilistic graphical model \cite{eryilmaz2016training}  and image learning \cite{Chen2014,chen2015memristor}. 
%and   accelerating       optimization solvers \cite{ren2017algorithm}.

\iffalse
Therefore,   memristor technology has been adopted to  implement   artificial neural network algorithms, e.g., 
back-propagation or other gradient-based algorithms, to accelerate its training process \cite{hasan15training,hasan2016reconfigurable,hasan2017chip,gokmen2016acceleration,agarwal2016resistive,li2014training,soudry2015memristor,yakopcic2017flexible,liu2016harmonica,wang2017group,ni2017energy}. 
 The computational merit  of memristor crossbars 
has also been stressed in other AI applications, such as dictionary  learning, sparse coding, and image processing  \cite{sheridan2017sparse,yu2015chip,seo2015chip,chen2015mitigating,chekad15,kadxu14,liu2017ultra,eryilmaz2016training,Chen2014,chen2015memristor}. 
\fi

Although   memristor-inspired  AI  applications  are  different, 
the common underlying theme is the   design 
of a mathematical programming solver for an optimization problem specified by a machine learning or data processing task.
Examples include 
 linear programming   for portfolio optimization \cite{mansini2014twenty}, nonlinear programming  for regression/classification \cite{bishop2006pattern}, and regularized   optimization  for sparse learning  \cite{bacjenmai12}. 
Therefore, a general question to be answered in this context  is:  \textit{how can one design a general memristor-based computation framework to accelerate the optimization procedure?} 
%This is the main focus of this paper. 

 %%% mapping difficulty
 %There exist challenges to map   optimization algorithms to memristor-based neuromorphic hardware.
The interior-point algorithm is one of the most commonly-used  optimization approaches implemented in software. It  begins at an interior point within the feasible region,   then applies a projective transformation so that the current interior point is the center of projective space, and then moves in the direction of the steepest descent \cite{boyd2004convex}. 
However, 
%there exist challenges to map  theinterior-point algorithm  to memristor-based neuromorphic hardware.
%To design the memristor-based optimization solver, 
\textcolor{black}{the inherent hardware limitations  prevent  the direct mapping from   the interior-point algorithm to memristor crossbars.} 
First,
a memristor crossbar only allows square matrices with nonnegative entries during computation, since
the memristance is always nonnegative. Second, the memristor crossbar suffers  from  hardware variations, which
degrade the reading/writing accuracy of  memristor crossbars.
 To circumvent the first difficulty, additional memristors were used to     represent negative elements of a square matrix \cite{hu2012hardware,kadetotad2014neurophysics,cai2016low}. 
 In particular, the work \cite{cai2016low} presented a memristor-based linear solver using the interior-point algorithm, which, however, requires programming of the  resistance state of memristors   at every iteration. % since the associated coefficient matrix  is udpated at each iteration of the interior-point algorithm. 
%This introduces the second difficulty, that is, 
Consequently, the linear solver in \cite{cai2016low} is prone to suffer from   hardware variations. 
Therefore, to successfully design memristor-based optimization solvers, it is crucial to co-optimize algorithm,
device and architecture so that the advantages of memristors can be fully utilized and the design  complexity and the non-ideal hardware effects can be minimized.
%Therefore, a more suitable optimization framework is desired that simplifies the configuration complexity onto memristor crossbars.
Our previous work \cite{ren2017algorithm,liu2017ultra} showed that the
alternating direction method of multipliers (ADMM) algorithm can take advantage of the hardware implementation of memristor crossbars.  
  % ADMM, an operator splitting method, is  a general   tool  to solve optimization problems, ranging from  linear program, quadratic program, to   semidefinite program \cite{o2013conic,boyparchupeleck11}.
  With the aid of ADMM, 
 one can decompose a complex problem   
 into subproblems that require
  matrix-vector multiplications and solution of systems of linear equations. The decomposed operations are  more easily mapped onto   memristor crossbars. \textcolor{black}{In this paper, we discuss  how to use the idea of ADMM to design memristor-based optimization solvers for solving  linear programs, quadratic programs and sparse optimization problems}. Different from the interior-point algorithm,  memristor crossbars are  programmed only  once, namely, independent of ADMM iterations. Therefore, the proposed memristor-based optimization framework is of highly resilient  to random noise and process variations.

In addition to designing a memristor-based optimization solver, we also discuss the application of memristors to  solve   eigenvalue problems. It is worth mentioning that computation of  eigenvalues/eigenvectors   is the key step in many AI applications and optimization problems, e.g.,  low-dimensional manifold learning \cite{fodor2002survey},  and semidefinite projection in semidefinite programming \cite{boyparchupeleck11}. In this paper, we present the generalization of the power iteration (PI) method using memristor crossbars. Conventionally, PI  only converges when the dominant eigenvalue is unique. Here, we adopt  
the Gram-Schmidt procedure   \cite{golub2012matrix}  to   handle convergence issues in the presence  of
  multiple dominant eigenvalues.  
We anticipate that this paper will inspire proliferation of memristor-based technologies, and fully utilize its extraordinary potential in emerging AI applications.

 The rest of the paper is organized as follows.  In Section\,\ref{sec: mem_concept},  we  review the memristor  technology for solving systems of linear equations. 
 In Section\,\ref{sec: mem_AI_opt}, we 
 %provide   a comprehensive survey on memristor-based AI applications, and 
 discuss  the idea of ADMM for convex optimization. In Section\,\ref{sec: LPQP}, we derive   memristor-based solvers for linear and quadratic programming. 
 In Section\,\ref{sec: sparse}, we apply the memristor technology for sparse optimization.  In Section\,\ref{sec: PCA}, we extend PI using memristors for eigenvalue/eigenvector computation. 
 In Section\,\ref{sec: Diss}, we 
 summarize  the topics presented in the paper and 
 and     discuss  future  research directions.

%\section{Background on Memristor Technology}
\section{Memristors in Solving Systems of Linear Equations}
\label{sec: mem_concept}

 % Memristor was introduced by L.O. Chua as the fourth element of circuit and was discovered by Hewlett-Packard  in 2008 \cite{chua1971memristor,strsni08}.
 \textcolor{black}{
A memristor has the unique property of recording the  %historical 
profile of  excitations on the device. That is,  the state (memristance) of a memristor   changes only when a certain voltage higher than a threshold
 is applied at its two terminals. 
%Otherwise, the memristor behaves like a resistor. 
This memristive property makes it an ideal candidate for use as non-volatile memory \cite{diperchu09,jochaebo10}. 
  Physical memristors can be
fabricated in a high density grid, and
the resulting memristor crossbar structure is attractive for performing  matrix-vector operations 
 due to its  high degree of parallelism %and high computational efficiency 
\cite{yakopcic2017flexible}. We elaborate on the memristor technology in the following.
}
%In this section, we will provide a background on memristor network and demonstrate its utility in acceleration of         matrix-vector operations.

\begin{figure}[htb]
\centering
%\centerline{ 
%\begin{tabular}{c}
 \includegraphics[width=0.6\textwidth,height=!]{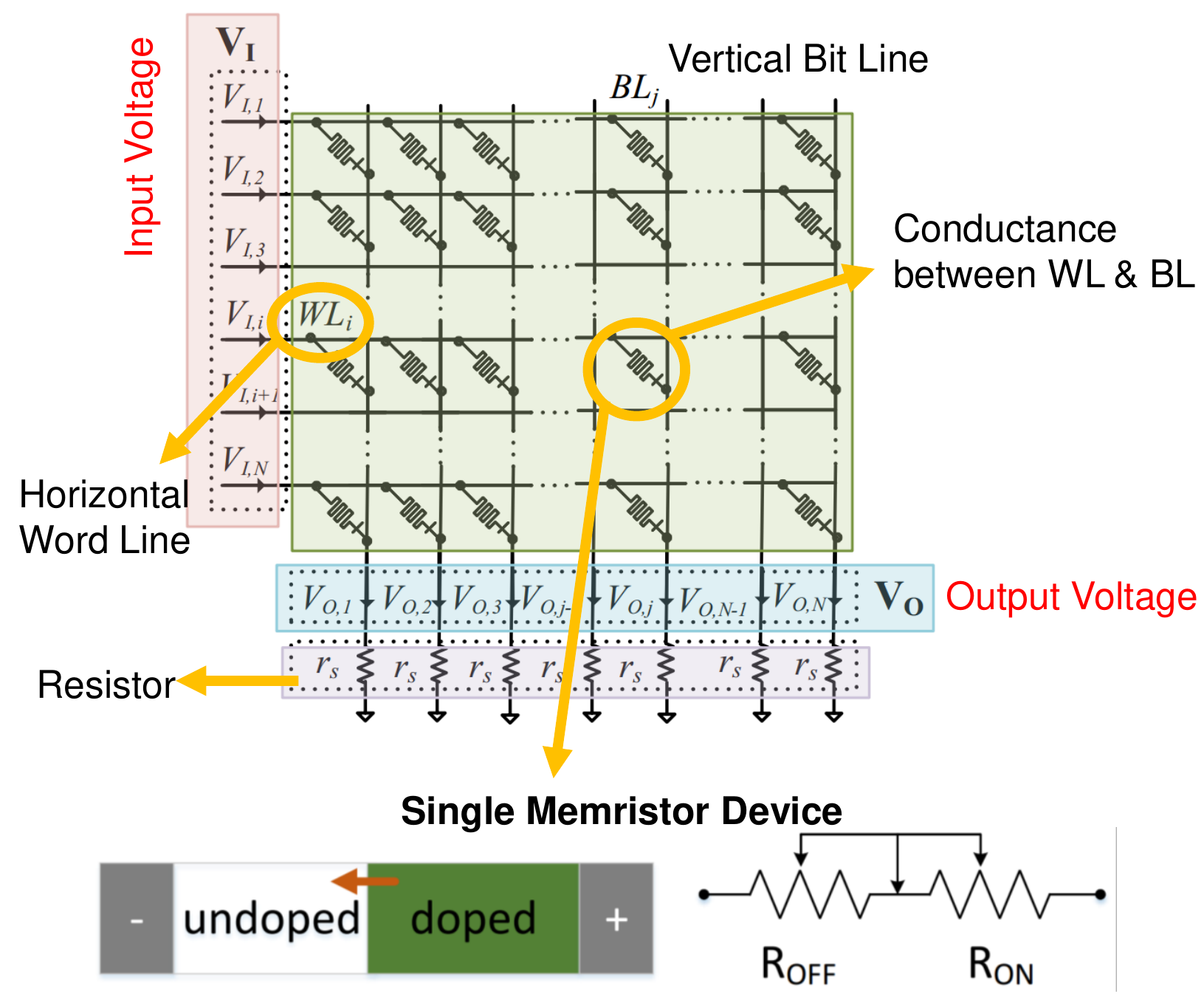}
%\end{tabular}}
\caption{Illustration of a memristor crossbar}
\label{fig: mem}
\end{figure}

%\subsection{Memristor-based matrix-vector operations}
A typical $N \times N$ memristor crossbar structure is illustrated in Fig.\,\ref{fig: mem}, where a memristor is connected between each pair of horizontal word-line (WL) and vertical bit-line (BL).  
This structure can be implemented with a small footprint, and each memristor can be re-programmed to different resistance states by  controlling the voltage of  WLs  and   BLs \cite{huli12cf,ni2017distributed,heittmann2012limits}.
Let $\mathbf V_{\mathrm{I}}$ denote a vector of input voltages on WLs. \textcolor{black}{We obtain the current at each BL  by measuring the voltage across a resistor with  conductance   $g_s$.} 
If the memristor at the connection between {WL}$_i$ and {BL}$_j$ has a conductance of $g_{i,j}$, then the   output voltage on the $j$th BL $ \mathbf V_{\mathrm{O},j}$ is  given by \cite{huli12cf}, 
\begin{align}
 \mathbf V_{\mathrm{O},j} = 
 \begin{bmatrix}
 \frac{g_{1,j}}{g_s + \sum_{i=1}^N g_{ij}} & \cdots &
  \frac{g_{N,j}}{g_s + \sum_{i=1}^N g_{ij}}
 \end{bmatrix}
 \mathbf V_{\mathrm{I}}, %\label{eq: VIO}
 \nonumber
\end{align}
or equivalently,
\begin{align}
 \mathbf V_{\mathrm{O}} = \mathbf C  \mathbf V_{\mathrm{I}},~
\mathbf C = 
\diag \left ( \left \{  \frac{1}{g_s + \sum_{i=1}^N g_{ij}}   \right \}_{j=1}^N \right ) 
\mathbf G^T, \label{eq: VIO}
\end{align}
where $\diag(\{ x_i \}_{i=1}^N)$ denotes a diagonal matrix with diagonal entries $x_1, x_2, \ldots, x_N$,  and $\mathbf G$ 
is the conductance matrix of  memristors  whose $(i,j)$th entry is given by $g_{i,j}$.
In \eqref{eq: VIO}, the desired coefficient matrix $\mathbf C$ 
%or its scaled version (within  memristors' conductance range \cite{yakopcic2017flexible})  
  can be realized by adjusting   memristor conductivities $\{ g_{i,j}\}$ and the bias resistor's  conductance $g_s$.  
 In order to avoid out-of-range coefficients in the memristor crossbar, a pre-scaling step is required to scale all matrix coefficients to fall into the memristors' conductance range. In this manner,
 one can perform  matrix-vector multiplications through   a pre-configured (or programmed) memristor crossbar.
   
 %The formula \eqref{eq: VIO} implies that one can realize  matrix-vector multiplications via adjusting the conductance matrix $\mathbf G$ of memristors.

%program   memristors in the crossbar based on a conductance matrix $\mathbf G$ obtained from $\mathbf C$ \cite{flocke2008fundamental,alibart2012high,yi2011feedback}, and conduct the operation of matrix-vector multiplication  by measuring    the output voltages of the programmed   crossbar. 

Reversing the above  operation, the memristor crossbar structure can also solve a  system of linear equations \cite{richter2015memristive}. Here, we assume that the solution exists and is unique. 
It is clear from \eqref{eq: VIO} that if we apply  $\mathbf V_{\mathrm{O}}$  on BLs, then  $\mathbf V_{\mathrm{I}}$ on  WLs becomes the solution of the linear system described by a  pre-configured memristor network. An appealing property of 
 the  memristor-based linear equation solver  is its high computational efficiency,
 an astonishing $O(1)$ time complexity   \cite{gokmen2016acceleration}, since
 the matrix-vector multiplication (or its reverse operation) is performed in a parallel fashion.
 %: input voltages  applied to all WLs (or BLs)  are multiplied by the conductance of the synaptic devices at each cross-point, leading to a weighted sum of voltages in each BL (or WL).
While this structure provides significant computational 
 advantages, there are  challenges   introduced by the  hardware restrictions of 
 memristors. 
First, in the linear system \eqref{eq: VIO},
only a non-negative coefficient matrix    can be mapped onto memristors. 
%Second, a memristor crossbar  is size-limited   (e.g., $1024 \times 1024$) due to manufacturing and performance considerations \cite{liu2015reno}.
%Third, 
Second,
a memristor crossbar suffers from   hardware variations that introduce computational errors while performing matrix-vector operations. 
In what follows, we elaborate on the aforementioned  challenges and present some possible solutions. 
%We elaborate on the aforementioned problems in the next subsection. 

%%% three aspects 
%%% writting coefficients nonnegative

%\subsection{Dealing with negative coefficients}
Since only non-negative coefficients can be mapped to  memristors, it is essential to design a general mechanism    that can deal with  negative   coefficients.  In previous  work \cite{li2014training,huli12cf}
   it has been  suggested that     negative numbers in a memristor system can be represented by  using two identical crossbars. Specifically, the weight matrix $\mathbf C$ is split into two parts $\mathbf C_1$ and $\mathbf C_{2}$ so that $\mathbf C = \mathbf C_1 - \mathbf C_2$, where $\mathbf C_1 = (\mathbf C)_+$,  $\mathbf C_2 = (-\mathbf C)_+$, and $(x)_+ = \max \{0, x \}$   is a positive operator taken elementwise   for a matrix argument. 
  Given nonnegative matrices $\mathbf C_1$ and $\mathbf C_2$, the matrix-vector multiplication \eqref{eq: VIO} can be obtained through  the subtraction   $\mathbf C_1 \mathbf V_{\mathrm{I}} - \mathbf C_2 \mathbf V_{\mathrm{I}}$ \cite{hu2012hardware,kadetotad2014neurophysics}. Instead of using two identical crossbars, we can
  eliminate the negative numbers by introducing auxiliary variables in the linear system  \eqref{eq: VIO}, 
   \begin{align}
    \mathbf V_{\mathrm{O}} = \mathbf C  \mathbf V_{\mathrm{I}} ~ \Longrightarrow
\begin{bmatrix}
( \mathbf C )_+ & \mathbf B \\
\mathbf D & \mathbf I_{\bar N}
\end{bmatrix}
\begin{bmatrix}
\mathbf V_{\mathrm{I}} \\
\bar {\mathbf V}_{\mathrm{I}}
\end{bmatrix} = \begin{bmatrix}
\mathbf V_{\mathrm{O}} \\
\mathbf 0_{\bar N}
\end{bmatrix},
%~ \Longleftrightarrow ~
%\tilde{\mathbf C}  \tilde {\mathbf z} = \tilde{\mathbf d},
\label{eq: VIO_ex}
\end{align} 
 where $\bar {\mathbf V}_{\mathrm{I}} \in \mathbb R^{\bar N}$ is a newly introduced variable, $\bar N$ is the number of nonzero columns of $(-\mathbf C)_+$ (namely, the number of columns of $\mathbf C$ that contain   negative elements), 
  $\mathbf B \in \mathbb R^{N \times \bar N}$ is formed by nonzero columns of $(-\mathbf C)_+$, 
 % of $(-\mathbf C)_+$ after its all-zero columns are removed,  
  $\mathbf D \in \mathbb R^{\bar N \times N}$  is
  a submatrix of  $\mathbf I_N$ whose row indices are given by column indices of nonzero columns of $(-\mathbf C)_+$,
 % is a submatrix of $\mathbf I_n$ 
  %after the   rows, indexed by the   columns of $\mathbf C$ that only contain nonnegative elements, are removed,  
  and $\mathbf 0_{\bar N}$ is a zero vector of size $\bar N$. 
  In \textcolor{black}{Table\,I}, we show that   \eqref{eq: VIO} can be recovered from \eqref{eq: VIO_ex} by eliminating    $\bar {\mathbf V}_{\mathrm{I}}$.   %\textcolor{red}{(Example in block; proof)}. 
We stress that compared to the use of an identical memristor crossbar (leading to $2N \times 2N$ memristor network), the proposed scheme \eqref{eq: VIO_ex} requires fewer  memristors, resulting in the memristor network of size $(N+ \bar N) \times (N + \bar N)$, where
$\bar{N} \leq N$.

\begin{framed}
\textbf{\textcolor{black}{Table I:}} Illustration of linear mapping \eqref{eq: VIO_ex}.
\begin{itemize}
\item $\mathbf C = (\mathbf C)_+ - (-\mathbf C)_+$, which yields $\mathbf V_{\mathrm{O}} = (\mathbf C)_+ \mathbf V_{\mathrm{I}} -   (-\mathbf C)_+ \mathbf V_{\mathrm{I}}$.
\item Let $\{ i_1, i_2, \ldots, i_{\bar N} \}$ denote the indices of nonzero columns of  $(-\mathbf C)_+$. Definitions of $\mathbf B = [\mathbf b_1, \ldots, \mathbf b_{\bar N}]$ and $\mathbf D = [\mathbf e_{i_1}, \ldots, \mathbf e_{i_N}]^T$ in \eqref{eq: VIO_ex} give
\[
(-\mathbf C)_+ = \sum_{j=1}^{\bar N} \mathbf b_j \mathbf e_{i_j}^T = \mathbf B \mathbf D.
\]
\item $\mathbf V_{\mathrm{O}} = (\mathbf C)_+ \mathbf V_{\mathrm{I}} -   \mathbf B \mathbf D \mathbf V_{\mathrm{I}} $ $\Longrightarrow $ 
$
\mathbf V_{\mathrm{O}} = (\mathbf C)_+ \mathbf V_{\mathrm{I}} + \mathbf B \bar {\mathbf V}_{\mathrm{I}}
$ with 
$
 \bar {\mathbf V}_{\mathrm{I}} =- \mathbf D \mathbf V_{\mathrm{I}}
$, which yields \eqref{eq: VIO_ex}. 
\end{itemize}
 
\end{framed}

 Moreover,  
 parameters of a memristor  crossbar may differ from the target values   due to variability in the fabrication process, environmental noise, and signal fluctuations from power supplies and neighboring wires  \cite{jo2009high}. 
 %Therefore, some degree of stochastic behavior  exists when a write pulse is applied to a memristor \cite{yu2011investigating}.  
%To employ memristor crossbars in technological and computational applications, 
Several methods have been proposed to mitigate these  impairments in hardware \cite{alibart2012high,yakopcic2017flexible,chekad15,chen2015mitigating,liu2017ultra}. In \cite{alibart2012high,yakopcic2017flexible}, feedback
programming techniques were used to improve the writing accuracy in memristor crossbars. In \cite{chekad15}, a read peripheral circuitry that functions as an analog-to-digital converter  was used to eliminate analog distortions. %However, the digitization could   decrease  the operation accuracy. 
In 
\cite{chen2015mitigating},  
multiple memristors were introduced to  update a single weight.  This method statistically averages out the conductance variations   in both time and space. 
However, it requires more memristors   and higher communication overhead. In addition to circuit-level techniques \cite{alibart2012high,yakopcic2017flexible,chekad15,chen2015mitigating},  we will show that 
 the non-ideal effects caused by hardware variations can also be mitigated  by optimizing the algorithm  prior to mapping to   memristor crossbars. %\cite{liu2017ultra}. 
 %This suggests that  an algorithm-hardware co-design and co-optimization framework is desired to overcome  the inherent limitations of memristor crossbars. 

 %\section{Memristor-Based    AI Applications and Convex Optimization}
  \section{Convex Optimization and ADMM}
  
  Although memristor-based AI applications are different  such as sparse learning and dictionary learning  \cite{sheridan2017sparse,yu2015chip,seo2015chip,chen2015mitigating,chekad15,kadxu14,liu2017ultra}, the principle  of designing  memristor-based computation accelerators is the same, namely,  recognizing the optimization problem underlying  the learning task  and mapping the corresponding optimization algorithm onto a memristor network. 
In what follows, we provide some   background on  mathematical programming and  focus on a solver called alternating direction method of multipliers (ADMM).

 \label{sec: mem_AI_opt}

\subsection{Preliminaries on convex optimization and ADMM}
%In that follows, we provide a brief background on  mathematical programming, namely, optimization.
In general, an optimization problem can be cast as
\begin{align}
\begin{array}{ll}
\displaystyle \minimize_{\mathbf x} &   \displaystyle  f(\mathbf x), \\
\st & \mathbf x \in \mathcal X,
\end{array}
\label{eq: prob_cox}
\end{align}
where $\mathbf x \in \mathbb R^n$ is the optimization variable, $f(\cdot)$ denotes the cost function to be minimized, and $\mathcal X$ denotes a  constraint set. In this paper, we focus on the convex version of problem \eqref{eq: prob_cox}, where  $f(\cdot)$ is a convex function and $\mathcal X$ is a convex set \cite{boyd2004convex}.  
In convex programming,  a
local minimum   given by a stationary point of  \eqref{eq: prob_cox} implies the global optimality. Convex optimization forms the foundation of many AI applications \cite{bishop2006pattern}. 

There exist many algorithms   to solve convex optimization problems, such as gradient-type first-order methods \cite{DP_book}, and primal-dual interior-point (second-order) methods \cite{boyd2004convex}. 
Compared to the conventional optimization methods, ADMM
%an operator splitting method, called alternating direction method of multipliers (ADMM), 
has drawn great attention in the last ten years \cite{boyparchupeleck11,parboy13}. 
The main advantage of ADMM is that  it allows us to split the optimization problem into subproblems, each of which can be solved   efficiently and, in some cases, analytically.     

A standard  problem that is suitable for  the application of ADMM is given by
\begin{align}
\begin{array}{ll}
\displaystyle \minimize_{\mathbf x, \mathbf y} &   \displaystyle  f(\mathbf x) + g(\mathbf y) \\
\st & \mathbf A \mathbf x + \mathbf B \mathbf y + \mathbf c = \mathbf 0,
\end{array}
\label{eq: prob_admm}
\end{align}
where $\mathbf x \in \mathbb R^n $ and $\mathbf y \in \mathbb R^m $ are optimization variables, $f(\cdot)$ and $g(\cdot)$ are convex functions, and $\mathbf A \in \mathbb R^{l \times n}$,   $\mathbf B \in \mathbb R^{l \times m}$, and $\mathbf c \in \mathbb R^l$ are appropriate coefficients associated with a  system of $l$  linear equality constraints.
 Problem \eqref{eq: prob_admm} reduces to problem \eqref{eq: prob_cox} when $\mathbf A = \mathbf I_n$, $\mathbf B = - \mathbf I_m$, $\mathbf c = \mathbf 0_l$, and $g(\cdot)$ is an indicator function on the convex set $\mathcal X$, namely,
 \begin{align}\label{eq: indicator_g}
 g(\mathbf y) = \left \{
 \begin{array}{ll}
 0 & \text{if $\mathbf y \in \mathcal X$} \\
 \infty & \text{otherwise}.
 \end{array}
 \right.
 \end{align}
  Here    $\mathbf I_n$ denotes the $n \times n$ identity matrix, and $\mathbf 0_n$ is the $n \times 1$ vector of all zeros.   
In what follows, while referring to identity matrices and vectors of all ones (or zeros), their
dimensions are  omitted for simplicity but can be inferred
from the context.   ADMM   is an iterative algorithm, and its $k$th iteration is given by \cite{boyparchupeleck11}
\begin{align}
&  \mathbf x^{k+1} = \displaystyle  
\argmin_{\mathbf x} \left \{ f(\mathbf x) +  (\boldsymbol \mu^k)^T (\mathbf A \mathbf x + \mathbf B \mathbf y^k + \mathbf c) + \frac{\rho}{2} \left \| \mathbf A \mathbf x + \mathbf B \mathbf y^k  + \mathbf c \right \|_2^2  \right \}
\label{eq: xstep}\\
&  \mathbf y^{k+1}   = \displaystyle \argmin_{\mathbf y} \left \{ g(\mathbf y) +  (\boldsymbol \mu^k)^T (\mathbf A \mathbf x^{k+1} + \mathbf B \mathbf y  + \mathbf c) + \frac{\rho}{2} \left \| \mathbf A \mathbf x^{k+1} + \mathbf B \mathbf y  + \mathbf c \right \|_2^2  \right \} \label{eq: wustep}\\
&  \boldsymbol \mu^{k+1} =  \boldsymbol \mu^k  + \rho(\mathbf A \mathbf x^{k+1} + \mathbf B \mathbf y^{k+1} + \mathbf c), \label{eq: dual} 
%\label{eq: dualstep_cs}
\end{align}
 where  $ \boldsymbol \mu$ is the Lagrangian multiplier (also known as the dual variable), $\rho$ is a positive   weight to penalize
 the augmented term associated with the equality constraint of \eqref{eq: prob_admm}, and $\| \cdot \|_2$ denotes the $\ell_2$ norm. The ADMM algorithm terminates when  an $\epsilon$-accuracy is achieved, namely, $\| \mathbf x^k - \mathbf y^k \|_2 \leq \epsilon$, and $\| \mathbf x^k - \mathbf x^{k-1} \|_2 \leq \epsilon$.
  ADMM has a linear convergence rate $O(1/K)$ for general convex optimization problems  \cite{heyuan12}, where $K$ is the number of iterations. In other words, given the stopping tolerance $\epsilon$, 
ADMM requires $O(1/\epsilon)$ iterations to converge. We remark that ADMM has a faster convergence rate than the  gradient decent algorithm, which has the convergence rate of   $ O(1/\sqrt{K})$. 
%In practice,  it is often the case that  ADMM converges   faster to   modest accuracy that is sufficient for many applications \cite{parboy13,boyparchupeleck11,liukarfarvar15,liufarmasvar14,linfarjov11}. 
 %To successfully  design   memristor-based optimization solvers,   it is crucial to  co-optimize algorithm, device and architecture so that the advantages of memristors can be fully utilized and   the designing complexity and  the non-ideal    hardware effects can be minimized.
In the next section,  we will show that ADMM   provides
a suitable framework for mapping to a memristor network.

%, where the key step    is to extract subproblems  in terms of   systems  of linear equations from ADMM. Based on \eqref{eq: VIO_ex}, the  linear equations can then  be efficiently  solved using memristor crossbars with astonishing $O(1)$ time complexity.
 
%Such principle drives both the selection of algorithms and the design evolution from CPU to CMOS application-specific integrated circuits (ASIC) to parallel architecture with resistive crosspoint array (PARCA) that we propose.

 \section{Memristor-Based  Linear and Quadratic Optimization Solvers}
 \label{sec: LPQP}
 
 In this section, 
 %we introduce  the idea of algorithm-hardware co-optimization for solving  
 we employ memristor crossbars to solve
 linear and quadratic programs.  Linear programs (LPs) and  quadratic programs (QPs)  are the most common optimization problems that are encountered in many applications  such as
resource scheduling, intelligent transportation,    portfolio optimization, smart grid and signal processing 
  \cite{aronofsky1964growing, dahrouj2010coordinated,liu2015sparsity,zhang2011optimal}. 
  The  interior-point  algorithm is a standard method  to solve
LPs  as well as QPs \cite{boyd2004convex}, with $O(n^3\sim n^{3.5})$ time complexity \cite{nemirovski2004interior}, where  $n$ is the number of optimization variables.    
The conventional interior-point  algorithm running on CPUs/GPUs has low degree of parallelism. 
By contrast, as we next demonstrate,  ADMM %is more suitable for memristor implementation
  breaks up optimization problems into subproblems involving the solution of linear equations, which lend themselves to the use of memristors for efficient computation.

%due to the ability to decompose the problem into independent subproblems that require solution of systems of linear  equations. 
%By contrast, memristor introduces a possible design resolution. %from CPU to parallel architecture with resistive crosspoint arrays, and the ADMM framework paves the way to implement optimization algorithms onto memristor crossbar arrays.  

%Such principle drives both the selection of algorithms and the design evolution from CPU to CMOS application-specific integrated circuits (ASIC) to parallel architecture with resistive crosspoint array (PARCA) that we propose.

\subsection{Linear optimization with memristors}

The standard form of LP is  expressed as follows,
 \begin{align}
 \begin{array}{ll}
 \displaystyle \minimize_{\mathbf x} & \mathbf d^T \mathbf x \\
 \st & \mathbf G \mathbf x = \mathbf h, \quad \mathbf x \geq \mathbf 0,
 \end{array}
 \label{eq: LP}
 \end{align}
 where $\mathbf x \in \mathbb R^n$ is the optimization variable, $\mathbf d \in \mathbb R^n$, $\mathbf G \in \mathbb R^{l \times n}$ and $\mathbf h \in \mathbb R^{l}$ are given parameters, and the last inequality constraint represents the elementwise inequalities $x_i \geq 0$ for $i = 1,2,\ldots, n$. In this paper, we assume that $\mathbf G$ is   of full   row rank.
 
%A memristor crossbar-based accelerator was proposed by using the PDIP algorithm in \cite{cai2016low}. Compared to PDIP,
%However, the principle of algorithm-hardware co-optimization drives a wiser strategy of algorithm selection and hardware design. 
%we show that ADMM provides a more efficient way to utilize memristors.  

We  begin by reformulating  problem \eqref{eq: LP} as the canonical form \eqref{eq: prob_admm} that is amenable to  the use of ADMM algorithm,
\begin{align}
 \begin{array}{ll}
 \displaystyle \minimize_{\mathbf x, \mathbf y} & \mathbf d^T \mathbf x + p(\mathbf x) + g(\mathbf y) \\
 \st &  \mathbf x = \mathbf y,
 \end{array}
 \label{eq: LP_admm}
 \end{align}
where $\mathbf y \in \mathbb R^n$ is a newly introduced optimization variable, and similar to 
\eqref{eq: indicator_g}, 
$p$ and $g$ are indicator functions, 
 with respect to  constraint sets
$\{\mathbf x \, | \, \mathbf G \mathbf x = \mathbf h \}$ and 
$\{\mathbf y \, | \, \mathbf y \geq \mathbf 0\}$, respectively. 
If we set $f(\mathbf x) = \mathbf d^T \mathbf x + p(\mathbf x)$, $\mathbf A = \mathbf I$,
$\mathbf B = -\mathbf I$ and $\mathbf c = \mathbf 0$, then problem \eqref{eq: LP_admm} is the same as  problem \eqref{eq: prob_admm}.

Based on   \eqref{eq: LP_admm}, the ADMM steps \eqref{eq: xstep}-\eqref{eq: dual}   become
\begin{align}
&  \mathbf x^{k+1} = \displaystyle  
\argmin_{\mathbf x} \left \{ \mathbf d^T \mathbf x + p(\mathbf x) + ( \boldsymbol \mu^k )^T (\mathbf x - \mathbf y^k) + \frac{\rho}{2} \left \|\mathbf x - \mathbf y^k \right \|_2^2  \right \}
\label{eq: xstep_LP}\\
&  \mathbf y^{k+1}   = \displaystyle \argmin_{\mathbf y} \left \{ g(\mathbf y) +  (\boldsymbol \mu^k)^T (\mathbf x^{k+1} - \mathbf y  ) + \frac{\rho}{2} \left \| \mathbf x^{k+1} - \mathbf y \right \|_2^2  \right \} \label{eq: wustep_LP}\\
&  \boldsymbol \mu^{k+1} =  \boldsymbol \mu^k  + \rho( \mathbf x^{k+1} - \mathbf y^{k+1} ). \label{eq: dual_LP} 
%\label{eq: dualstep_cs}
\end{align}
As we show next, the primary advantage of employing ADMM here is that problem \eqref{eq: xstep_LP} can be readily solved using   memristor crossbars, and  problem \eqref{eq: wustep_LP} yields a closed-form solution that 
only involves elementary vector operations.
 
Problem \eqref{eq: xstep_LP} is equivalent to 
\begin{align}
 \begin{array}{ll}
 \displaystyle \minimize_{\mathbf x} & %\mathbf d^T \mathbf x +
 \displaystyle  \frac{\rho}{2}\|  \mathbf x - \boldsymbol \alpha \|_2^2 \\
 \st &  \mathbf G \mathbf x = \mathbf h,
 \end{array}\label{eq: xstep_LP_equi}
\end{align} 
where $\boldsymbol \alpha \Def \mathbf y^k - (1/\rho) (\boldsymbol \mu^k + \mathbf d)$. The solution of problem \eqref{eq: xstep_LP_equi} is determined by its Karush-Kuhn-Tucker (KKT) conditions \cite{boyd2004convex},
$
 \rho (\mathbf x - \boldsymbol \alpha) + \mathbf G^T \boldsymbol \lambda = \mathbf 0$, and $\mathbf G \mathbf x = \mathbf h$,
where $ \boldsymbol \lambda \in \mathbb R^l$   is the Lagrangian multiplier.  
The KKT conditions  imply a system of linear equations 
\begin{align}\label{eq: lin_sys_LP}
\mathbf C \begin{bmatrix}
\mathbf x \\
 \boldsymbol \lambda
\end{bmatrix}
  = \begin{bmatrix}
\rho \boldsymbol \alpha  \\
\mathbf h
\end{bmatrix},\quad  \mathbf C = \begin{bmatrix}
\rho \mathbf I & \mathbf G^T \\
\mathbf G & \mathbf 0
\end{bmatrix}.
\end{align}
Based on \eqref{eq: VIO_ex}, the linear system \eqref{eq: lin_sys_LP} can be efficiently mapped to memristor crossbars  
  by configuring their memristance values according to the matrix $\mathbf C$. %\textcolor{red}{[Give a remark on the structure of $\mathbf C$]}
  
On the other hand, problem \eqref{eq: wustep_LP} is equivalent to
  \begin{align}
 \begin{array}{ll}
 \displaystyle \minimize_{\mathbf y} &  \displaystyle \frac{\rho}{2}\|  \mathbf y - \boldsymbol \beta \|_2^2 \\
 \st &  \mathbf y \geq \mathbf 0,
 \end{array}\label{eq: ystep_LP_equi}
\end{align}
where $\boldsymbol \beta \Def \mathbf x^{k+1} + (1/\rho) \boldsymbol \mu^k$. The solution of problem \eqref{eq: ystep_LP_equi} is determined by the projection of $\boldsymbol \beta$ onto the nonnegative orthant,
\begin{align}
\mathbf y^{k+1} = (\boldsymbol \beta)_+. \label{eq: lp_y}
\end{align}
Note that the positive part operator   $(\cdot)_+ $ in \eqref{eq: lp_y} can be  readily  implemented using elementary logical or digital operations. 

\begin{figure}[htb]
\centering
%\centerline{ 
%\begin{tabular}{c}
 \includegraphics[width=0.5\textwidth,height=!]{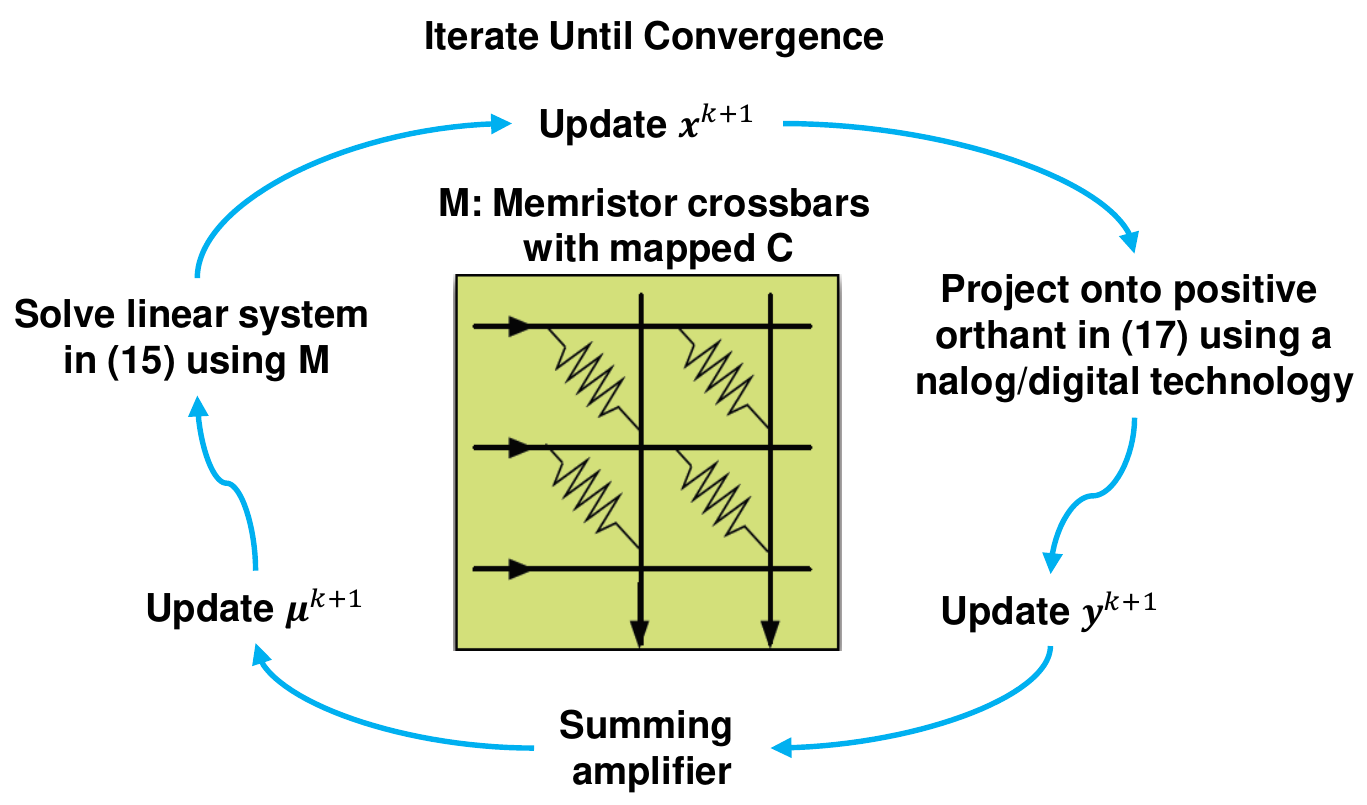}
%\end{tabular}}
\caption{Memristor crossbar based  solution framework in linear programming.}
\label{fig: LP}
\end{figure}

We summarize the memristor-based LP solver in Fig.\,\ref{fig: LP}. Although LP is a relatively simple optimization problem, the LP solver  illustrates our general idea and paves the way 
for numerous memristor-based applications in optimization problems.
Our solution framework  offers two major advantages. First, 
in the linear system \eqref{eq: lin_sys_LP},
the coefficient matrix $\mathbf C$ is independent of the ADMM iteration so that memristors need to  be configured only once. This feature makes it more attractive than   gradient-type and interior-point algorithms, where  memristors have to be reconfigured at each iteration \cite{chen2015mitigating}. Second, ADMM splits a complex problem into  subproblems, each of which is   easier  to solve and implement in  hardware.

\begin{figure}[htb]
\centerline{ 
\begin{tabular}{cc}
 \includegraphics[width=0.25\textwidth,height=!]{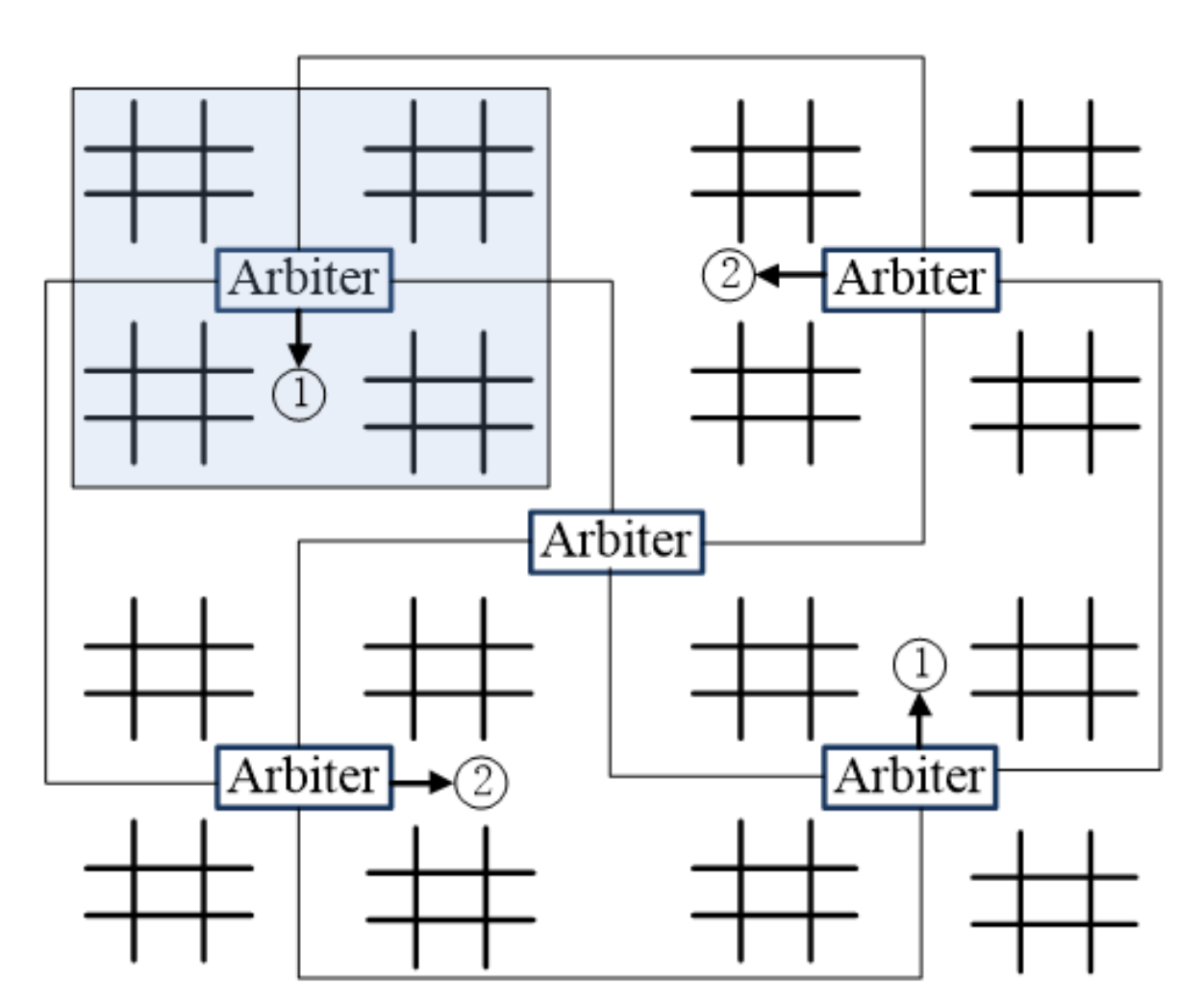}
& \includegraphics[width=0.25\textwidth,height=!]{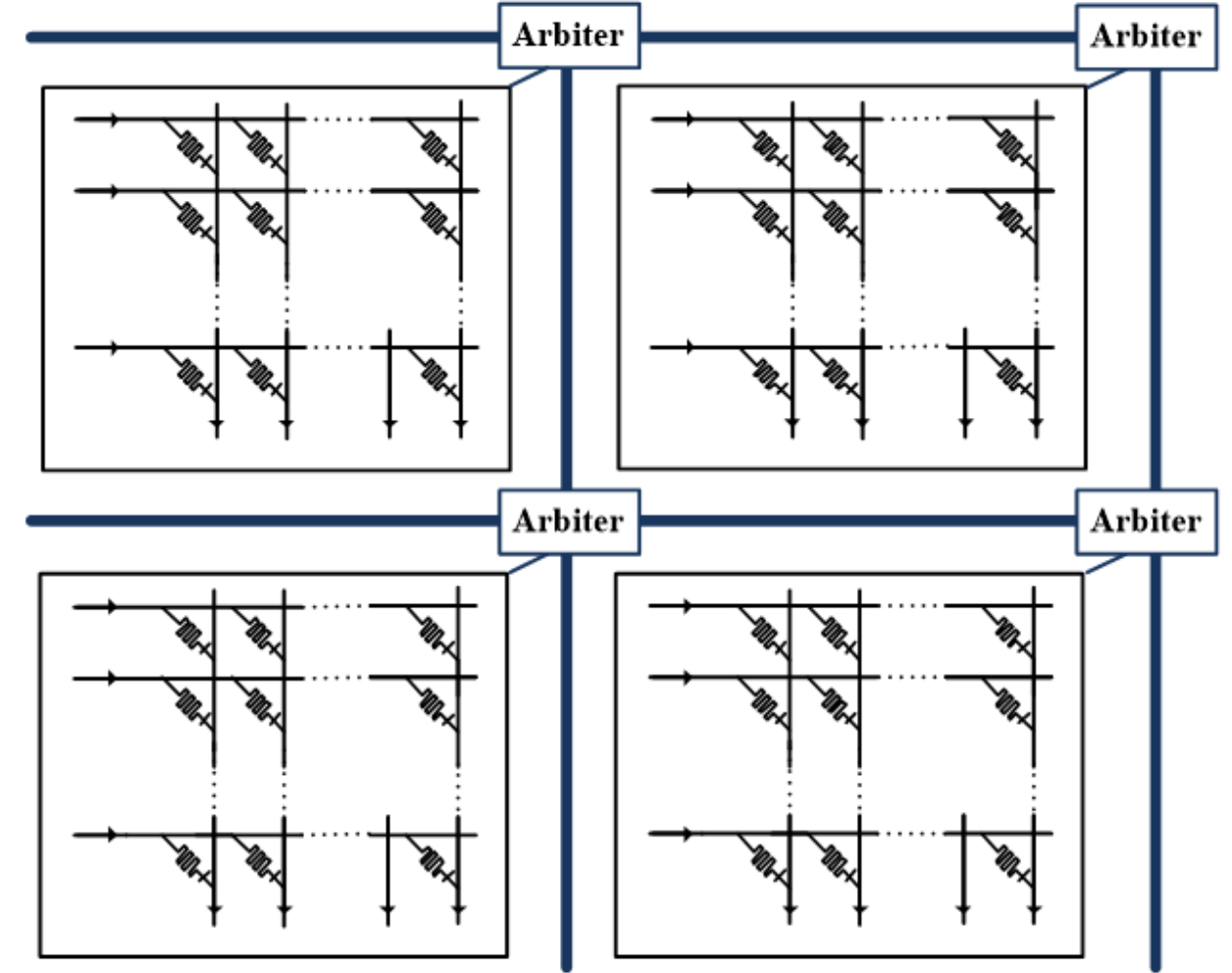}\\
(a) & (b)
\end{tabular}}
\caption{Examples of NoC structures coordinating multiple memristor crossbars. (a) Four crossbar arrays are grouped and controlled by one arbiter. The resulting higher-level group is controlled by a higher-level arbiter. (b) Mesh network-based structure of memristor crossbars.}
\label{fig: memcom}
\end{figure}

\textcolor{black}{
We remark that a memristor crossbar  is size-limited   (e.g., $1024 \times 1024$) due to manufacturing and performance considerations \cite{liu2015reno}.
To improve its scalability,  analog network-on-chip (NoC) communication structures can be adopted to   effectively coordinate multiple memristor crossbars for supporting large-scale applications
 \cite{liu2015reno,yakopcic2015hybrid,liu2016harmonica,dally2001route}. Data transfers within the NoC structure maintain analog form and are managed by the NoC arbiters.
Two potential analog NoC structures for multiple memristor crossbars are presented in Fig.\,\ref{fig: memcom}. 
Fig.\,\ref{fig: memcom}(a) shows 
a   hierarchical structure of memristor crossbars \cite{liu2015reno}, where four crossbar arrays are grouped and controlled by one arbiter, and those groups again form a higher-level group controlled by a higher-level arbiter.   Fig.\,\ref{fig: memcom}(b) shows  a mesh network-based structure of memristor crossbars, which resembles a mesh network-based NoC structure in multi-core systems \cite{dally2001route}. 
}
%In addition to the NoC structure, it is also possible to  reduce the problem size 
%by simplifying the problem
%prior to 
%hardware implementation \cite{liu2017ultra,cai2016low,ren2017algorithm} . 
%algorithm-level techniques  could be used to simplify matrix-vector operations prior to hardware implementation.
%For example, the work \cite{liu2017ultra,cai2016low,ren2017algorithm}   employed an alternating optimization method to  
%split the large-scale problem into small-scale subproblems, which are then mapped to memristor crossbars. 
%solve these subproblems in an iterative manner. Similar applications 

%examples include memristor-based   solver  for linear programming and dictionary learning problems  \cite{cai2016low,kadetotad2014neurophysics}.  

\subsection{Quadratic optimization with memristors}
QP is an optimization problem whose objective  and constraint functions involve quadratic and/or linear terms. 
There exist many variants of QP, such as a second-order cone program (SOCP) and a quadratically constrained quadratic program (QCQP) \cite{boyd2004convex}.  
In this section, we focus on the design of a memristor-based solver for SOCP, since   it is possible to convert a QCQP into a SOCP, e.g., homogeneous QCQP that excludes linear terms \cite{ren2017algorithm}.

SOCP is a convex program for minimizing a linear cost function subject to linear  and second-order cone constraints,
\begin{align}\label{eq: socp}
\begin{array}{ll}
 \displaystyle \minimize_{\mathbf x} & \mathbf d^T \mathbf x  \\
 \st &  \mathbf G \mathbf x = \mathbf h, \quad \| \mathbf x_{1:(n-1)} \|_2 \leq x_n,
\end{array}
\end{align}
where $\mathbf x \in \mathbb R^n$ is the optimization variable, $\mathbf G$ and $\mathbf h$ are given parameters, 
$\mathbf x_{1:(n-1)} $ denotes a vector that consists of the first $n-1$ entries of $\mathbf x$, and $x_n$ is the $n$th entry of $\mathbf x$.
The last constraint in \eqref{eq: socp} is known as the second-order cone constraint. 

Similar to \eqref{eq: LP_admm}, we can rewrite problem \eqref{eq: socp} in the   canonical form \eqref{eq: prob_admm} that is  amenable to the ADMM algorithm
\begin{align}\label{eq: socp_admm}
\begin{array}{ll}
 \displaystyle \minimize_{\mathbf x, \mathbf y} & \mathbf d^T \mathbf x + p(\mathbf x) + g(\mathbf y) \\
 \st &  \mathbf x = \mathbf y,
\end{array}
\end{align}
where $\mathbf y \in \mathbb R^n$ is the newly introduced optimization variable, and
$p$ and $g$ are indicator functions  with respect to  constraint sets
$\{\mathbf x \, |  \, \mathbf G \mathbf x = \mathbf h \}$ and 
$\{\mathbf y \, | \, \| \mathbf y_{1:(n-1)} \|_2 \leq y_n \}$, respectively.

\begin{figure}[htb]
\centerline{ 
\begin{tabular}{c}
 \includegraphics[width=0.7\textwidth,height=!]{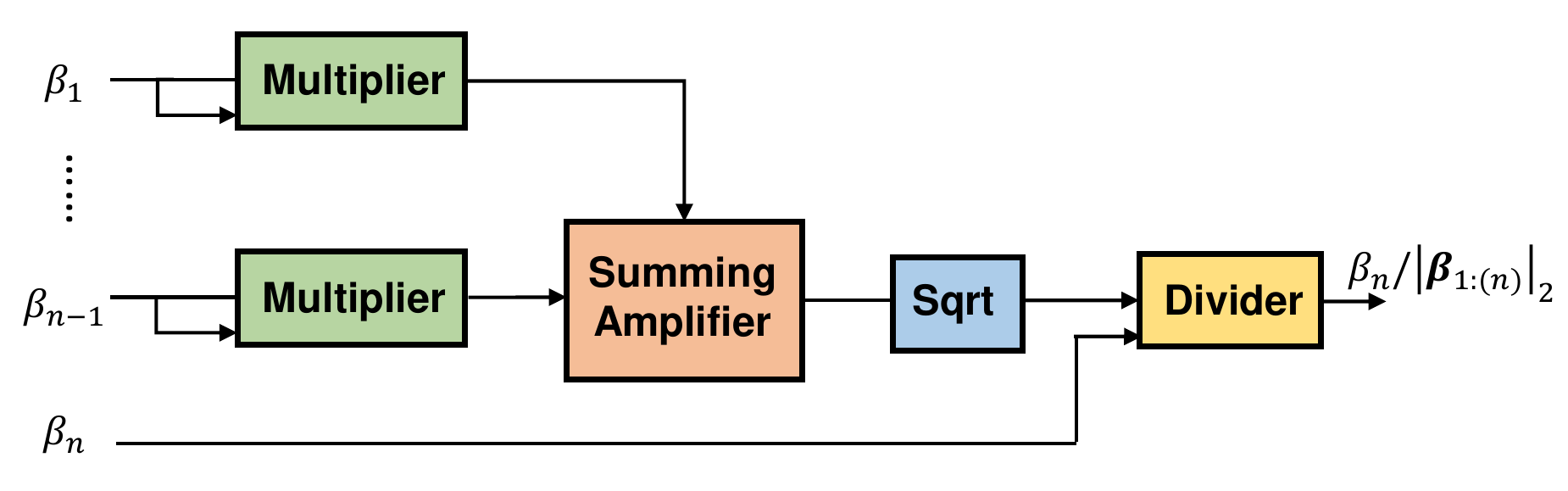} 
\end{tabular}}
\caption{Schematic illustration of the hardware system for calculating $\mathbf y^{k+1}$ in \eqref{eq: prox_socp}.}
\label{fig: vec_norm_hard}
\end{figure}

Following  \eqref{eq: xstep}-\eqref{eq: dual}, the ADMM algorithm for solving problem \eqref{eq: socp_admm} includes   subproblem \eqref{eq: xstep_LP_equi} with respect to the variable $\mathbf x$, step \eqref{eq: dual_LP}  for updating dual variables $\boldsymbol \mu$, and   a specific $\mathbf y$-minimization problem \eqref{eq: wustep}, 
  \begin{align}
 \begin{array}{ll}
 \displaystyle \minimize_{\mathbf y} &  \displaystyle  \frac{\rho}{2}\|  \mathbf y - \boldsymbol \beta \|_2^2 \\
 \st &  \| \mathbf y_{1:(n-1)} \|_2 \leq y_n,
 \end{array}\label{eq: ystep_socp}
\end{align}
where recall from \eqref{eq: ystep_LP_equi} that  $\boldsymbol \beta = \mathbf x^{k+1} + (1/\rho) \boldsymbol \mu^k$. The solution of problem \eqref{eq: ystep_socp} is given by  projecting $\boldsymbol \beta $ onto a second-order cone \cite{parboy13},
\begin{align}\label{eq: prox_socp}
\mathbf y^{k+1} = \left \{
\begin{array}{ll}
0 & \| \boldsymbol \beta_{1:(n-1)} \|_2 \leq - \beta_n \\
\boldsymbol \beta & \| \boldsymbol \beta_{1:(n-1)} \|_2 \leq  \beta_n \\
\frac{1}{2}\left (1+\frac{\beta_n}{\| \boldsymbol \beta_{1:(n-1)} \|_2} \right ) \left [ \boldsymbol \beta_{1:(n-1)}^T,\| \boldsymbol \beta_{1:(n-1)} \|_2 \right ]^T  &\| \boldsymbol \beta_{1:(n-1)} \|_2 \geq | \beta_n |.
\end{array}
\right.
\end{align}
Similar to the memristor-based LP solver, the ADMM step \eqref{eq: xstep} reduces to the solution of  a system of linear equations that can be mapped onto memristor crossbars. 
In the ADMM step \eqref{eq: ystep_socp},  we can use peripheral circuits including analog multipliers and summing amplifiers  to evaluate the  vector  norm in \eqref{eq: prox_socp} \cite{huli13,wenwu15}; see schematic illustration in Fig\,\ref{fig: vec_norm_hard}. %Or we can convert the vector  to the digital domain and then calculate its  norm. 

To summarize, one may  exploit the alternating structure of ADMM to design   memristor-based optimization solvers.  The crucial property to enable this is that  ADMM helps in  extracting parallel operations of  
matrix/vector multiplication/addition which can    be  implemented using
memristor crossbars and  
  elementary hardware elements.

  \subsection{Performance evaluation}
  
  In what follows, we present  empirical results that show 
  the effectiveness of the proposed  memristor-based optimization framework to solve LPs and QPs\footnote{All the codes  will be made public once the paper is accepted.}.  Since the presence
of hardware variations leads to a reduced configuration accuracy on memristor crossbars, the matrix $\mathbf C$ in \eqref{eq: lin_sys_LP} is 
 actually modified to $\tilde{\mathbf C} = \mathbf C + \boldsymbol \Sigma$, where $\boldsymbol \Sigma$ denotes a random matrix whose elements are  i.i.d.   zero-mean Gaussian random variables. The quantity $\| \boldsymbol \Sigma \|_F/\| \mathbf  C \|_F$ then provides the level of hardware variations, where $\| \cdot \|_F$ denotes the Frobenius norm of a matrix. 
 %We demonstrate both the convergence and the accuracy of  our proposed solution framework against the level of hardware variation. 
In the presence of hardware variations, we  compare the  solution $\mathbf x$ above  to the  optimal solution ${\mathbf x}^*$ obtained from the off-the-shelf interior-point solver CVX \cite{cvx}, that excludes the effect of hardware variation.  We adopt
$\| \mathbf x - {\mathbf x}^{*} \|_2/\| {\mathbf x}^{*} \|_2$
(averaged over $50$ random trials)  to measure the error  between $\mathbf x$ and  $\mathbf x^*$. 
In ADMM,  the augmented  parameter and the stopping tolerance are set to be  $\rho \in \{ 0.1, 1, 10, 100 \}$ and $\epsilon = 10^{-3}$.

\begin{figure}[htb]
\centerline{ 
\begin{tabular}{cc}
 \includegraphics[width=0.5\textwidth,height=!]{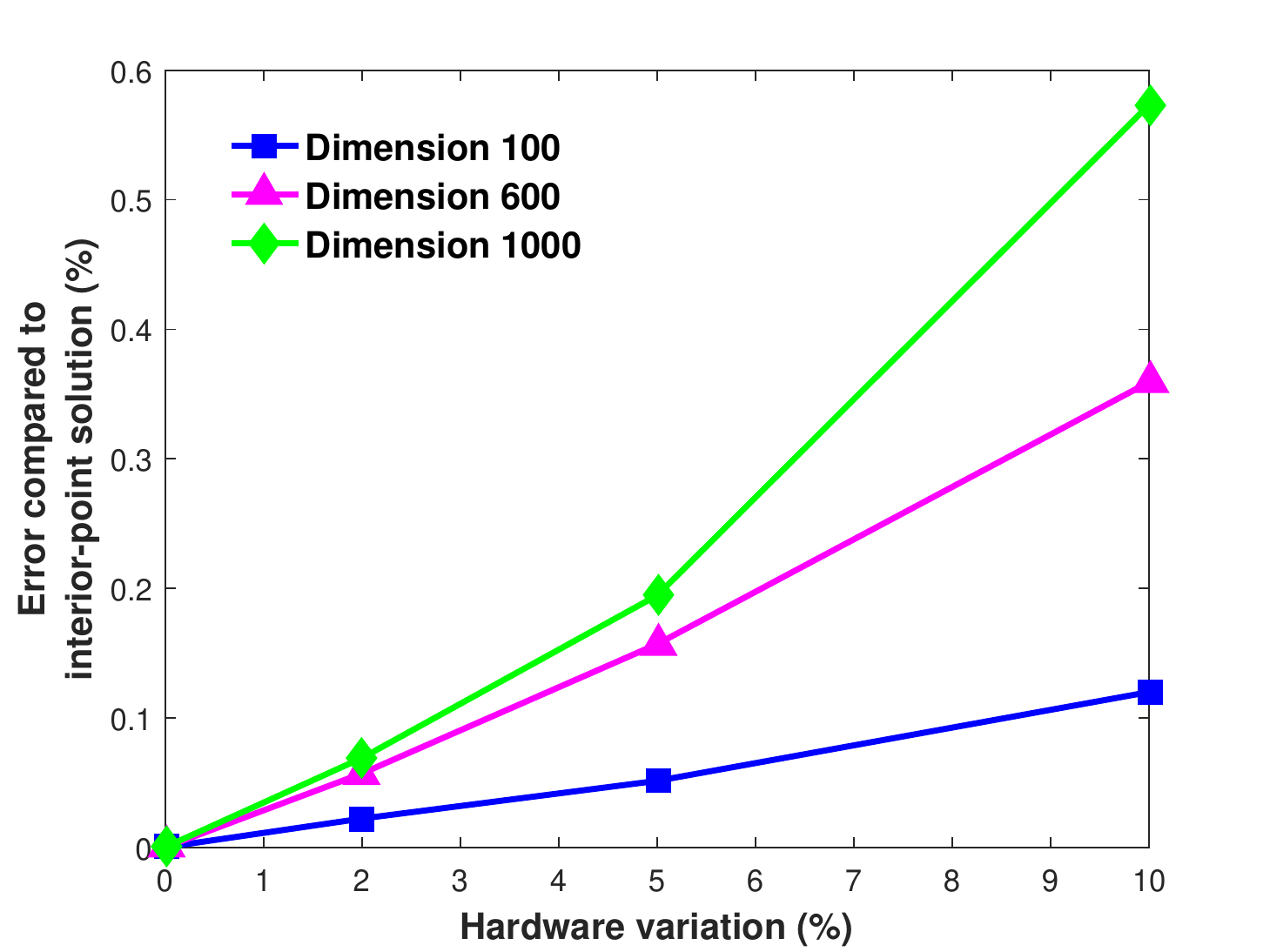}
& \includegraphics[width=0.5\textwidth,height=!]{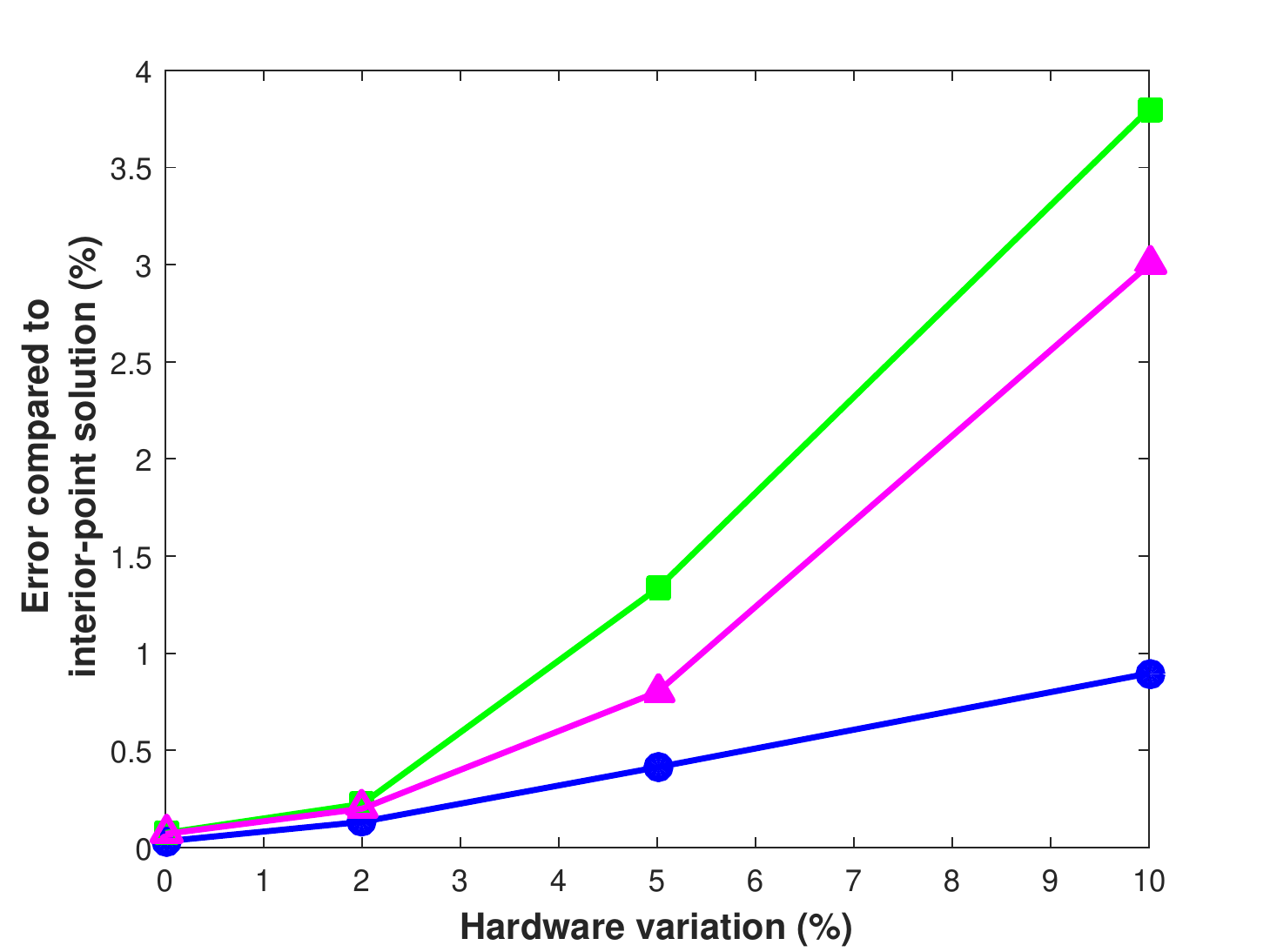}\\
(a) & (b)
\end{tabular}}
\caption{Solution accuracy versus level of hardware variations for different problem sizes $n \in \{100, 600, 1000 \}$. (a) Memristor-based  LP solver. (b) Memristor-based  QP solver with the same legend as (a).}
\label{fig: LP_QP_error}
\end{figure}

\begin{figure}[htb]
\centerline{ 
\begin{tabular}{cc}
 \includegraphics[width=0.5\textwidth,height=!]{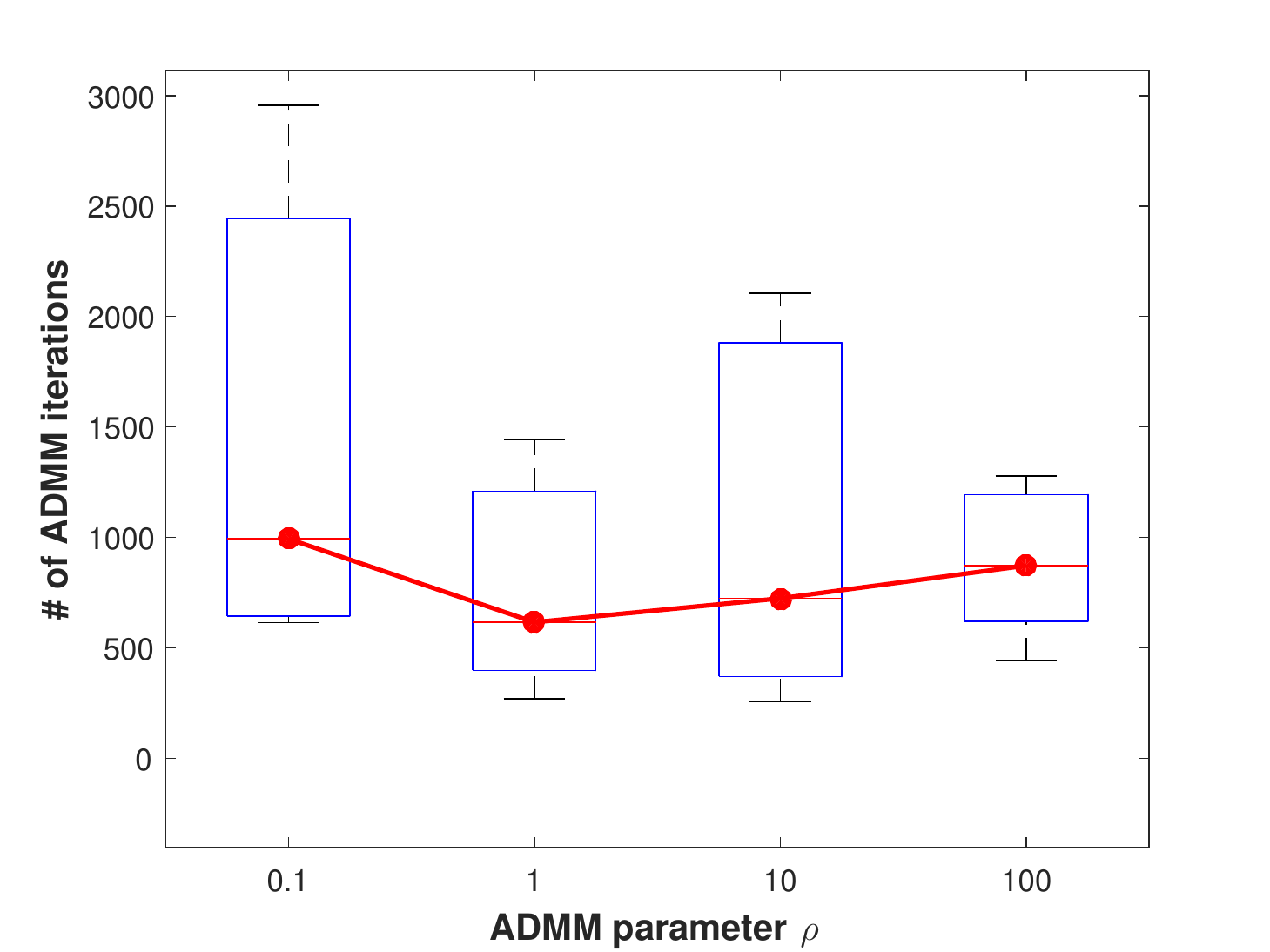}
& \includegraphics[width=0.5\textwidth,height=!]{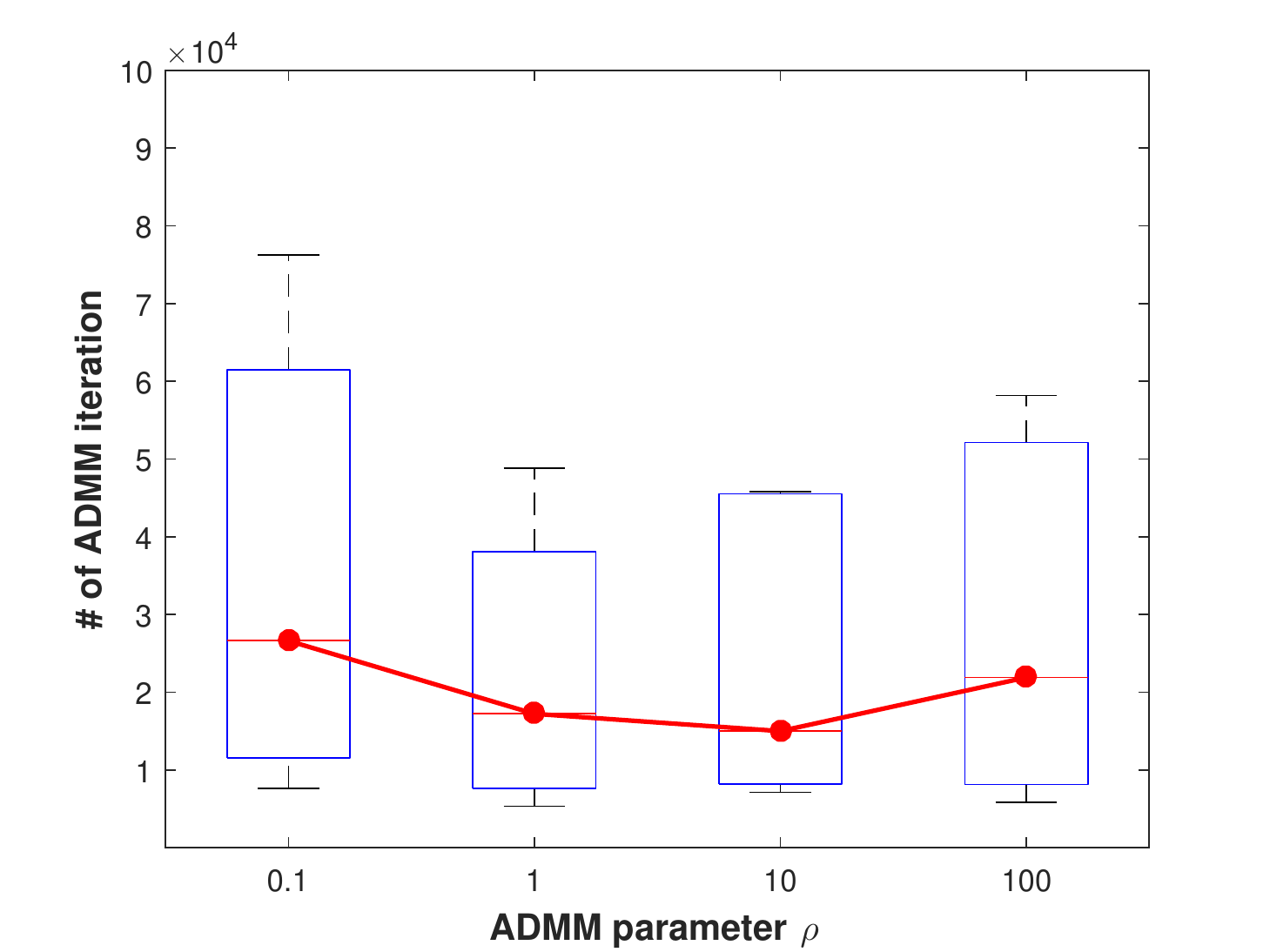}\\
(a) & (b)
\end{tabular}}
\caption{Number  of ADMM iterations to obtain an  $\epsilon$-accuracy solution for different values of ADMM parameter $\rho$ under $10\%$ hardware variation. (a) Memristor-based LP solver; (b) Memristor-based QP solver.}
\label{fig: LP_QP_iter}
\end{figure}

 In Fig.\,\ref{fig: LP_QP_error}, we present the difference between the memristor-based  solution and the variation-free interior-point solution  as a function of  the level of hardware variations for problems with dimension $n \in \{ 100,600, 1000\}$. When the hardware  variation is excluded, 
the memristor-based  solution yields the same accuracy as  the interior-point solution. 
As the problem size or the hardware variation increases, the difference from the interior-point solution increases. However, the induced error is  always below $5\%$. 
In Fig.\,\ref{fig: LP_QP_iter}, we further show the convergence of the memristor-based  solution framework as a function of  the choice of the ADMM parameter $\rho$. For each value of   $\rho$,  
 $50$ random trials were performed, each of which involved   $10\%$ hardware variation. 
We find that the convergence of the memristor-based  approach (to achieve $\epsilon$-accuracy solution)   is robust to   hardware variations and the choice of  ADMM parameter $\rho$. Compared to LP, QP requires more iterations to converge due to its higher complexity.  Moreover, a moderate choice of $\rho$, e.g., $\rho = 1$ in this example, improves the convergence  of the memristor-based  approach.

 \section{Memristor-Based Sparse Learning}
 \label{sec: sparse}
Sparse learning  is  concerned with the problem of finding intrinsic sparse patterns of  variables to be optimized. %e.g, signals to be recovered in compressive sensing, and linear regression coefficients in sparse regression.
This problem  is central to machine learning and big-data processing.  
%In the context of machine learning and big-data processing, 
Examples of applications include model selection in  regression/classification,  dictionary learning, matrix completion in recommendation systems,   image
restoration,  graphical modelling, natural language processing, resource management in sensor networks, and compressive sensing \cite{bacjenmai12,liu2016resource,slavakis2014modeling,chepuri2016sparse}. It is often the case that we can cast sparse learning as  an optimization problem that involves sparsity-inducing
regularizers, 
such as the $\ell_1$ norm, mixed $\ell_1$ and $\ell_2$ norms, and the nuclear norm
 \cite{bacjenmai12}. In this section, 
we focus on the problem of robust compressive sensing (CS),   which recovers sparse signals from noisy observations \cite{canwak08mag}. 
We remark that CS yields a problem formulation similar to LASSO \cite{tibshirani1996regression}, sparse coding \cite{sheridan2017sparse} and sensor selection problems \cite{liu2014sparsity}.
Previous research efforts \cite{qaibil13,canromtao06, dono06,canromtao06it,daimil09,xuhass07,canwak08mag} focused on  software-based approaches for sparse signal recovery,   with the support of   CPUs/GPUs. Here    we  discuss approaches to  employ memristor crossbars   to design   %ultra-efficient on-chip
CS solvers.

\subsection{Preliminaries on CS}
Let $\mathbf z_* \in \mathbb R^p$ be a sparse or compressible vector, e.g., a digital signal or image, to be recovered. We have access to  measurements 
$
\mathbf h = \mathbf H \mathbf z_* + \mathbf v$,
where $q \ll p$,
$\mathbf H \in \mathbb R^{q \times p}$ is a given measurement matrix, such as a random Gaussian matrix, and $\mathbf v \in \mathbb R^q $ is a stochastic or deterministic error   with bounded energy $\| \mathbf v\|_{2} \leq \xi$.
%Here we consider the high-dimentional case in which $l \ll n$. %and $\| \cdot \|_2$ denotes the $\ell_2$ norm of a vector.

The main goal of   CS is to stably recover the unknown sparse signal $\mathbf z_*$ from   noisy measurements $\mathbf h$. It has been shown in \cite{can06} that   stable recovery  can be achieved in polynomial time  by solving the convex optimization problem  for robust CS
\begin{align}
\begin{array}{ll}
\displaystyle \minimize_{\mathbf z} & \| \mathbf z \|_1 \\
\st & \|\mathbf H \mathbf z - \mathbf h \|_2 \leq \xi,
\end{array}
\label{eq: prob_CS}
\end{align}
where $\mathbf z \in \mathbb R^n$ is the optimization variable, and $\| \cdot \|_1$ denotes the $\ell_1$ norm of a vector. 
In  problem \eqref{eq: prob_CS}, the $\ell_1$ norm is introduced to promote the sparsity of   $\mathbf z$ \cite{canromtao06}. Note that problem \eqref{eq: prob_CS} can also be  formulated in the form of LASSO or sparse coding \cite{tibshirani1996regression,sheridan2017sparse}
\begin{align*}
\begin{array}{ll}
\displaystyle \minimize_{\mathbf z} & \|\mathbf H \mathbf z - \mathbf h \|_2^2 + \gamma \| \mathbf z \|_1,
\end{array}
%\label{eq: prob_CS_Lasso}
\end{align*}
where $\gamma$ is a regularization parameter that governs the tradeoff between the least square error and the sparsity of $\mathbf z$. In what follows, we focus on the problem formulation in \eqref{eq: prob_CS}. 

%We note that  problem \eqref{eq: prob_CS}  is a special instance of a second order cone program \cite{Boyd2004_bk}, which is typically solved by the interior-point algorithm \cite{canromtao06}. 

\subsection{Memristor-based accelerator for solving CS problems}
Similar to memristor-based linear and quadratic optimization solvers,
the key step to   successfully applying   memristor crossbar arrays to CS problems is  to extract   subproblems,  with the aid of ADMM,  that solve   systems of   linear equations. 
By introducing three new optimization variables $\mathbf s \in \mathbb R^q$, $\mathbf w \in \mathbb R^p$ and $\mathbf u \in \mathbb R^q$, 
problem \eqref{eq: prob_CS} can be reformulated in a way that lends itself to the application of ADMM,
\begin{align}
\begin{array}{ll}
\displaystyle \minimize & f(\mathbf z, \mathbf s)  + \| \mathbf w \|_1 + p(\mathbf u)  \\
\st &   \mathbf z - \mathbf w = \mathbf 0, \quad \mathbf s - \mathbf u = \mathbf 0,% \\
%&   \mathbf A \mathbf x -\mathbf u - \mathbf y = 0,
\end{array}
\label{eq: prob_CS_admm}
\end{align}
where $\mathbf z$, $\mathbf s $, $\mathbf w $ and $\mathbf u$ are optimization variables,  and $f$ and $p$ are    indicator functions  corresponding to the constraints 
of problem \eqref{eq: prob_CS}, namely, 
 \begin{align}
f(\mathbf z, \mathbf s) = \left \{
\begin{array}{ll}
0 &  \mathbf H \mathbf z - \mathbf s = \mathbf h  \\
\infty & \text{otherwise},
\end{array}
\right.
\label{eq: indicator2}
\end{align}
and
 \begin{align}
p(\mathbf u) = \left \{
\begin{array}{ll}
0 &  \| \mathbf u \|_2 \leq \xi \\
\infty & \text{otherwise}.
\end{array}
\right.
\label{eq: indicator1}
\end{align}
In \eqref{eq: prob_CS_admm},
\textcolor{black}{the introduction of new variables $\mathbf s$, $\mathbf w$ and $\mathbf u$ together with the indicator functions \eqref{eq: indicator2}\,--\,\eqref{eq: indicator1}  allows us to split the original constrained problem into subproblems for solving systems of linear equations, and elementary proximal operations related to   the $\ell_1$ norm and the Euclidean ball constraint \cite{parboy13}.}

We recall from the standard form of ADMM  given by \eqref{eq: prob_admm} that
if we set $\mathbf x = [\mathbf z^T, \mathbf s^T]^T$, $\mathbf y = [\mathbf w^T, \mathbf u^T]$, $g(\cdot) = \|  \cdot \|_1 + g^\prime (\cdot)$, $\mathbf A = \mathbf I$, $\mathbf B = -\mathbf I$ and $\mathbf c = \mathbf 0$, then problem \eqref{eq: prob_admm} reduces to the CS problem \eqref{eq: prob_CS_admm}.
As a result,
the ADMM step \eqref{eq: xstep} with respect to $\mathbf z$ and $\mathbf s$ can be written as
\begin{align}
\begin{array}{ll}
\displaystyle \minimize_{\mathbf z, \mathbf s}  &   \displaystyle \frac{\rho}{2}   \| \mathbf z - \boldsymbol \alpha_1  \|_2^2 +   \frac{\rho}{2}  \| \mathbf s - \boldsymbol \alpha_2 \|_2^2 \\
\st & \mathbf H \mathbf z - \mathbf s = \mathbf h,
\end{array}
\label{eq: xstep_cs_new}
\end{align}
where  $\boldsymbol \alpha_1 \Def \mathbf w^k - (1/\rho) \boldsymbol \mu_1^k$,    $\boldsymbol \alpha_2  \Def \mathbf u^k   - (1/\rho)  \boldsymbol \mu_2^k $, $\boldsymbol \mu = [\boldsymbol \mu_1^T, \boldsymbol \mu_2^T]^T \in \mathbb R^{p+q}$ is  the  vector of dual variables corresponding to problem \eqref{eq: prob_CS_admm}, 
and $k$ is the ADMM iteration number. 
The solution of problem \eqref{eq: xstep_cs_new} is given by KKT conditions:
$\rho  \mathbf z + \mathbf H^T \boldsymbol \lambda = \rho \boldsymbol \alpha_1  $, 
$
\rho \mathbf s -  \boldsymbol \lambda = \rho \boldsymbol \alpha_2
$, and $\mathbf H \mathbf z - \mathbf s = \mathbf h$, where $\boldsymbol \lambda \in \mathbb R^q$ is the Lagrangian multiplier corresponding to problem \eqref{eq: xstep_cs_new}. These form a system of linear equations
\begin{align}
\mathbf C
\begin{bmatrix}
\mathbf z \\
\mathbf s\\
\boldsymbol \lambda
\end{bmatrix} = 
\begin{bmatrix}
\rho \boldsymbol \alpha_1 \\
\rho \boldsymbol \alpha_2\\
\mathbf h
\end{bmatrix},
\quad 
\mathbf C = \begin{bmatrix}
\rho \mathbf I_p & \mathbf 0 & \mathbf H^T \\
\mathbf 0 & \rho \mathbf I_q & - \mathbf I _q\\
\mathbf H & - \mathbf I_q & \mathbf 0
\end{bmatrix}.
\label{eq: sol_x_cs}
\end{align}
Based on  \eqref{eq: VIO_ex},    
the linear system \eqref{eq: sol_x_cs} can be mapped onto a memristor network by 
configuring its memristance values. Recall that  a programmed memristor crossbar 
only requires a constant-time complexity $O(1)$  to solve problem \eqref{eq: sol_x_cs}.

The ADMM step \eqref{eq: wustep} with respect to $\mathbf w$ and $\mathbf u$  becomes
 \begin{align}
\begin{array}{ll}
\displaystyle \minimize_{\mathbf w, \mathbf u} & \displaystyle \| \mathbf w \|_1 + p (\mathbf u)  + \frac{\rho}{2}\|  \mathbf w - \boldsymbol \beta_1  \|_2^2 + \frac{\rho}{2}\| \mathbf u  -\boldsymbol \beta_2 \|_2^2,
\end{array}
\label{eq: wustep_cs_new}
\end{align}
where $\boldsymbol \beta_1 \Def \mathbf z^{k+1} + (1/\rho) \boldsymbol \mu_1^k$ and 
$\boldsymbol \beta_2 \Def \mathbf s^{k+1}  %\mathbf H \mathbf z^{k+1} - \mathbf h  
+ (1/\rho)  \boldsymbol \mu_2^k$.
Note that problem \eqref{eq: wustep_cs_new} can be decomposed  into two  problems with respect to $\mathbf w$ and $\mathbf u$:
\begin{align}\label{eq: sub_cs_l1}
\left \{
\begin{array}{l}
 \displaystyle  \minimize_{\mathbf w} ~  \| \mathbf w \|_1  + \frac{\rho}{2}  \|  \mathbf w - \boldsymbol \beta_1 \|_2^2, \\
  \displaystyle   \minimize_{\mathbf u} ~ \|  \mathbf u - \boldsymbol \beta_2 \|_2^2, ~ \st ~ \|  \mathbf u \|_2 \leq \xi.
\end{array}
\right.
\end{align}
Both problems in \eqref{eq: sub_cs_l1} can be solved analytically \cite{liu2017ultra}
\begin{align}\label{eq: sol_sub_cs_l1}
\left \{
\begin{array}{l}
 \mathbf w^{k+1} =   (\boldsymbol \beta_1 - 1/\rho \mathbf 1 )_{+} - (-\boldsymbol \beta_1 - 1/\rho \mathbf 1  )_+,  \\
\mathbf u^{k+1}  = \min\{ \xi, \| \boldsymbol \beta_2 \|_2 \} \frac{\boldsymbol \beta_2 }{\| \boldsymbol \beta_2\|_2}, 
\end{array}
\right.
\end{align}
where recall that $(\cdot )_+$ is the positive part operator. 
%Similar to ADMM steps in LPs and QPs, we can use elementary  hardware-based  operations to realized the positive part operator and the $\ell_2$ norm in \eqref{eq: sol_sub_cs_l1}.

 Similar to  LPs and QPs,
the hardware design of the   memristor-based  CS solver mainly consists of two parts. The first part is the memristor-based linear system solver, in which memristor crossbars are only programmed once since the coefficient matrix $\mathbf C$ in \eqref{eq: sol_x_cs} is independent of ADMM iterations. The second part is the digital or analog implementation of  the solution to problem \eqref{eq: sol_sub_cs_l1}. 
This requires the calculation of  the $\ell_2$ norm of a
vector that can be realized using elementary logic or digital operations; similar to Fig.\,\ref{fig: vec_norm_hard}.
%which can be realized by using elementary digital logics. %In addition, the updating   dual variable $\boldsymbol \mu^{k+1}$   involves standard addition/subtraction calculations in digital or analog domains. 
The ADMM-based solution   exhibits low hardware complexity.

%It is known from \cite{chekad15} that   memory devices   relying on the ion/defects motion typically show noticeable device variations from cell-to-cell and from cycle-to-cycle.   
%Therefore, to turn a memristor-based optimization solver  into   real-life applications, we face the following obstacle.
%The presence of hardware variations  leads to a reduced   reading/writing accuracy of  the coefficient matrix $\mathbf C$ that is configured in memristor crossbars.
We finally remark that  one can adjust the ADMM parameter $\rho$ to avoid the  hardware    variation-induced singularity for $\mathbf C$ in \eqref{eq: sol_x_cs}. This is supported by 
% in the presence of hardware variations,  the coefficient matrix $\mathbf C$ in \eqref{eq: sol_x_cs} may suffer  the  variation-induced singularity. However,  we can adjust the ADMM parameter $\rho$ so that $\mathbf C$ is  well-conditioned, supported by 
  the invertibility of the Schur complement of $\mathbf C$ \cite{petersen2008matrix},   $ (-1/\rho) (\mathbf I + \mathbf H \mathbf H^T) $. 
Specifically, if $\rho$ is too large,   the  Schur complement approaches zero (towards singularity). If $\rho$ is too small,   the effect of hardware variations on  $\mathbf H$ is magnified. %which leads to the poor optimization performance.  
Therefore, an appropriate choice of $\rho$ enhances 
the robustness of memristor-based optimization solvers   to  hardware variations.
%Therefore, an appropriate choice of $\rho$ enhances  the robustness of memristor-based optimization solvers   to device and process variations. 

 %\textcolor{red}{[Figures]}
 
 \subsection{Performance evaluation}
Next, we empirically show the effectiveness of the proposed solution framework for sparse signal recovery. Assume that the original signal $\mathbf z^*$ is of 
% Let  $\mathbf z^*$ be the   sparse signal to be recovered, with  
 dimension $p= 1024$  with $s \in \{ 10, 50, 100, 150,  200 \}$ nonzero elements. These nonzero spike positions are chosen randomly, and  their values are chosen independently   from the standard normal distribution.
To specify the CS problem  \eqref{eq: prob_CS},
  a  measurement matrix $\mathbf H \in \mathbb R^{500 \times 1024}$ with i.i.d. entries from   the standard normal distribution is generated, and  set $\xi = 10^{-3}$. The vector of   measurement noises $\mathbf v$ is drawn from the normal distribution $\mathcal N( \mathbf 0, 0.01 \mathbf I)$. 
To evaluate the recovery performance, the following two measures are employed a) the difference between the recovered signal $\mathbf z$ and the true sparse signal $\mathbf z^*$, namely, $\| \mathbf z - \mathbf z^* \|$, and b) the sparse pattern difference between   $\mathbf z$ and $\mathbf z^*$. 
%that yields the estimation accuracy of nonzero spike positions. 
All the performance measures are obtained by averaging  over $50$ random trials. For ADMM, unless specified otherwise, we set  $\rho  \in \{ 0.1, 1, 10, 100 \}$ and  $\epsilon = 10^{-3}$  for
its  augmented  parameter  and    stopping  tolerance.

\begin{figure}[htb]
\centerline{ 
\begin{tabular}{cc}
 \includegraphics[width=0.5\textwidth,height=!]{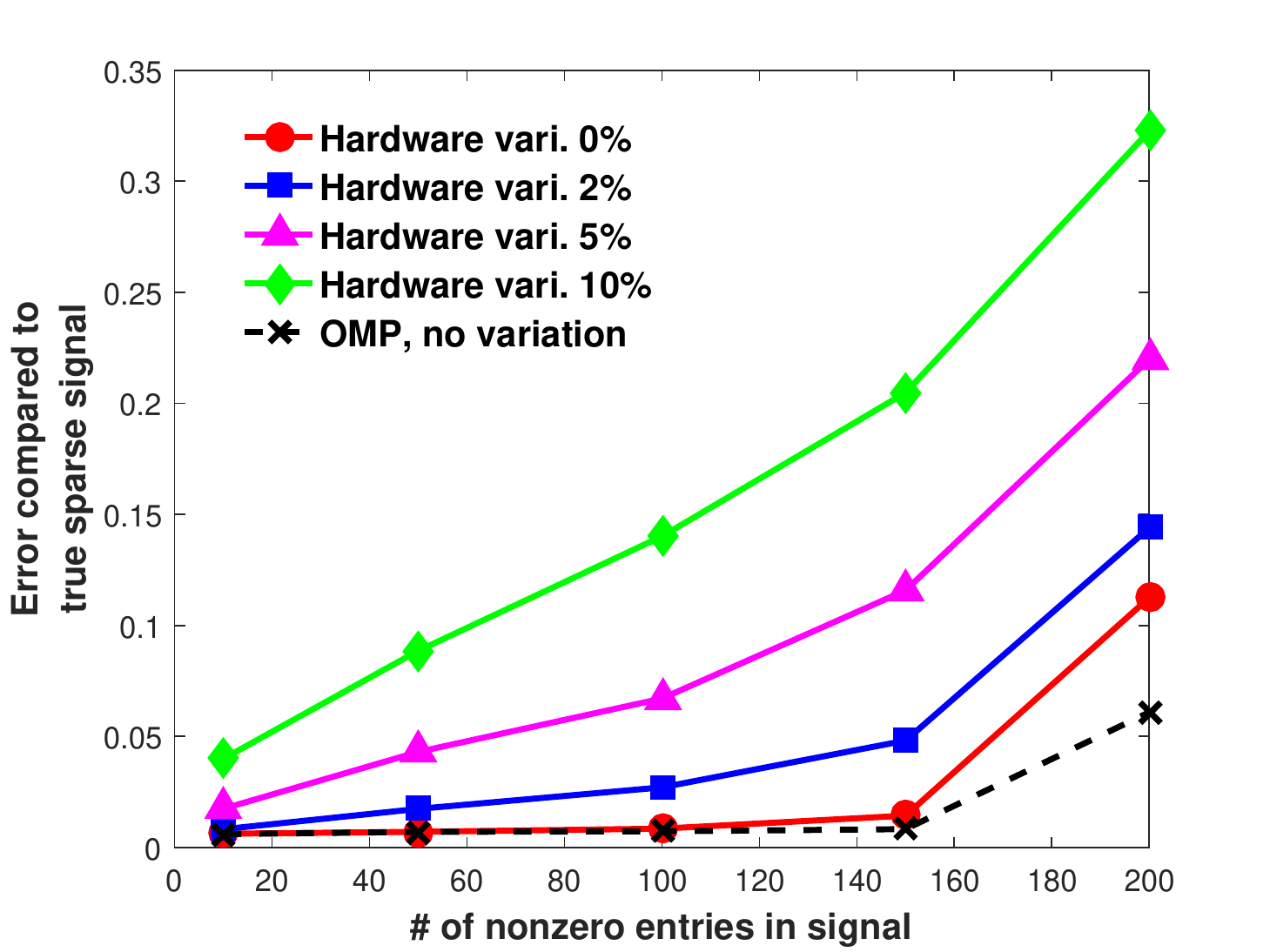}&
 \includegraphics[width=0.5\textwidth,height=!]{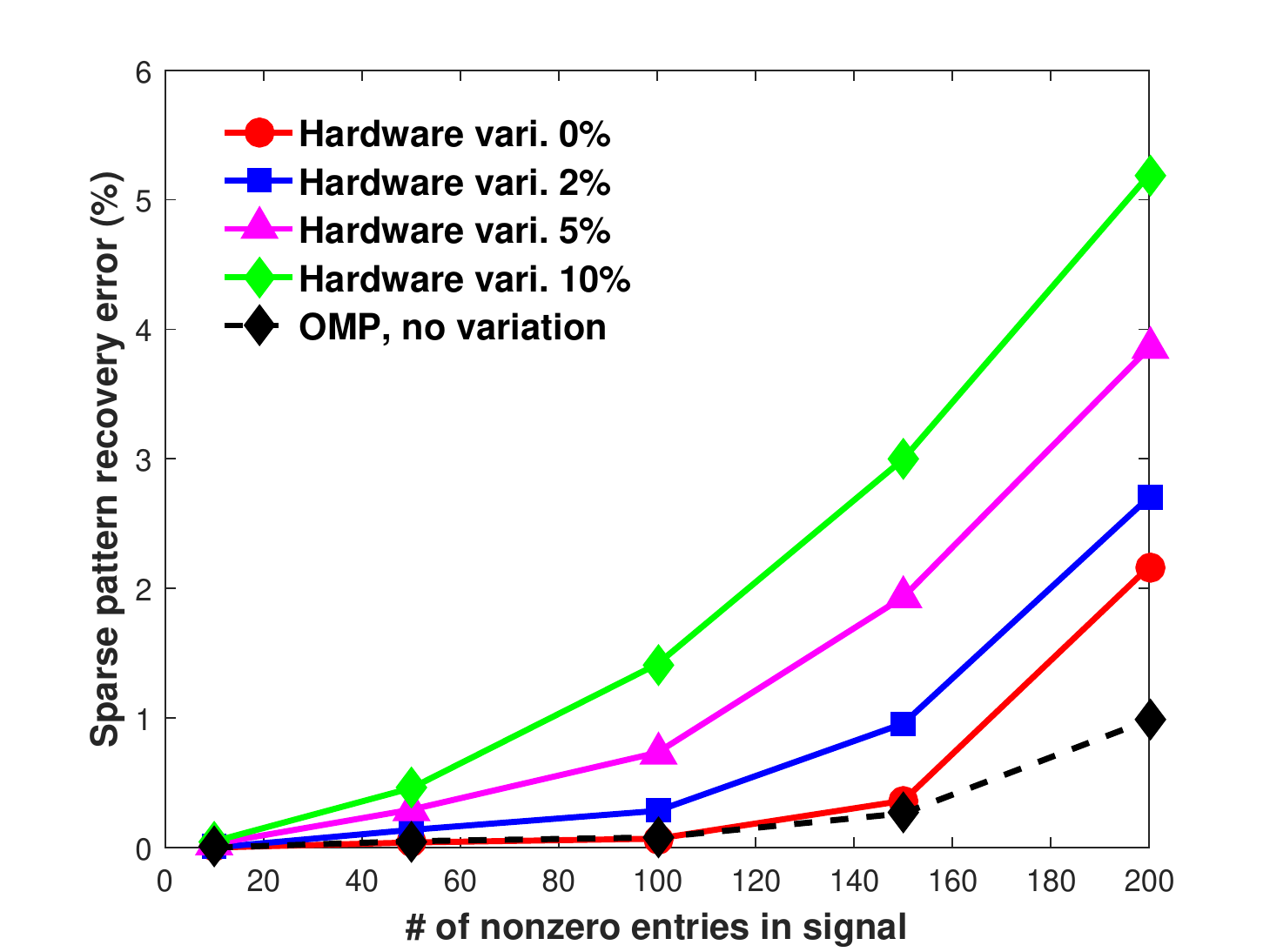}
\\
(a) & (b) \\
 \includegraphics[width=0.5\textwidth,height=!]{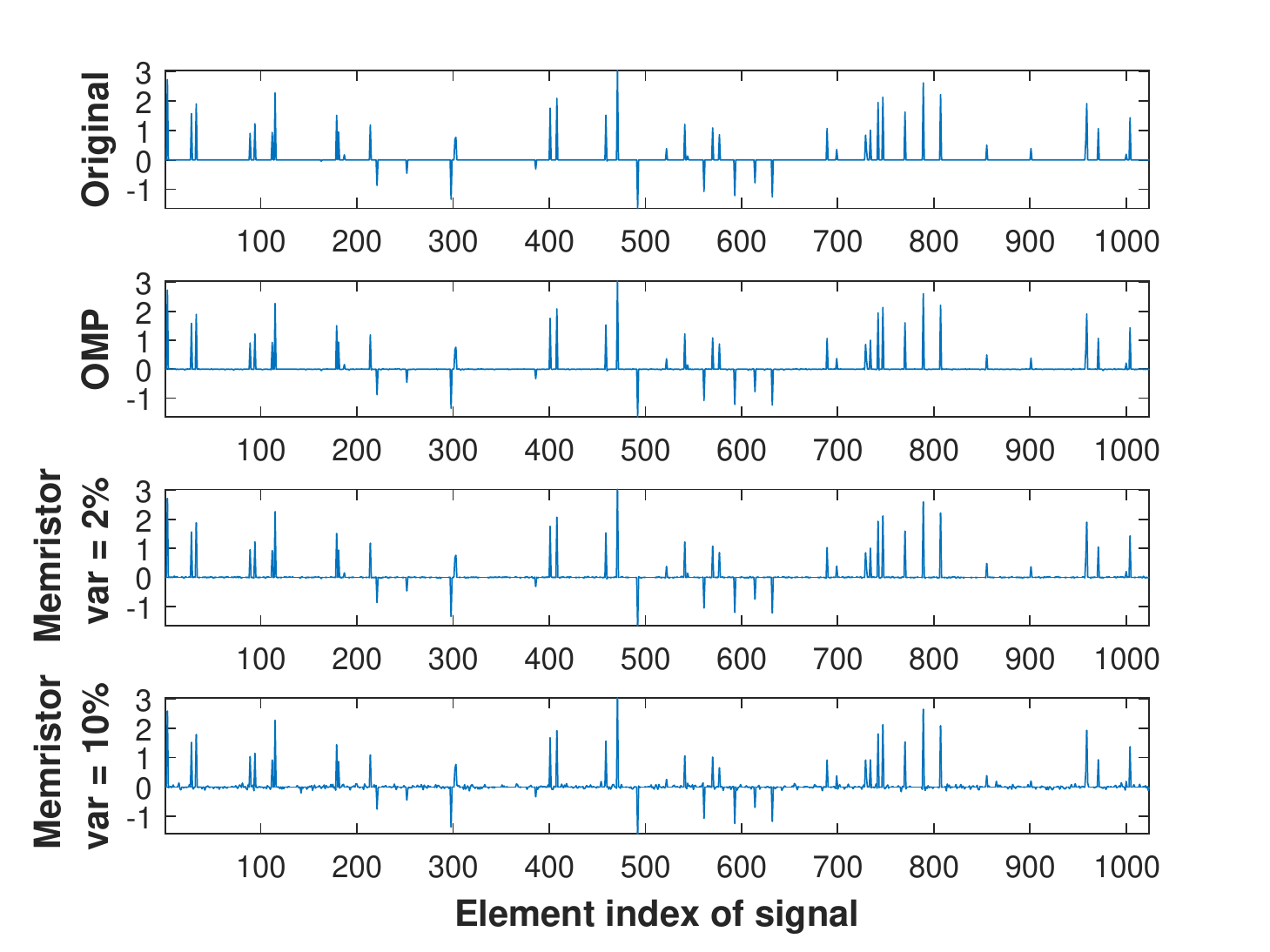}
 &  \includegraphics[width=0.5\textwidth,height=!]{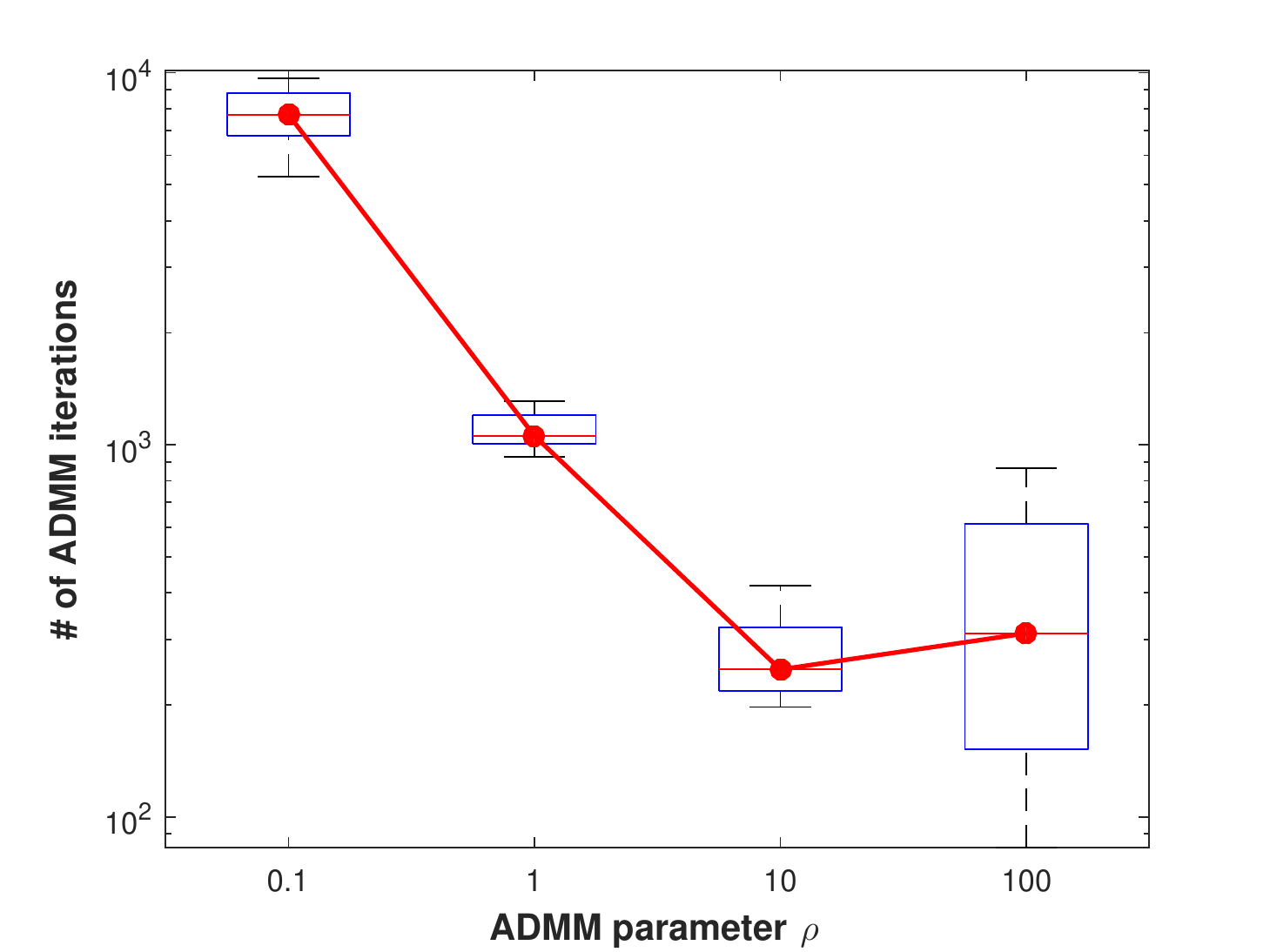}
 \\
 (c) & (d) 
\end{tabular}}
\caption{Sparse signal recovery performance under different levels of hardware variation. (a) Error with respect to  the true signal versus the sparsity level $s$. (b) Sparse pattern recovery  error versus the sparsity level $s$. (c) Recovered signals with $s = 50$ nonzero entries. (d) Number of iterations required for  convergence versus the ADMM parameter $\rho$.}
\label{fig: CS_err}
\end{figure}

 In Fig.\,\ref{fig: CS_err}, we present the performance of sparse signal recovery by using the memristor-based  solution framework. 
 Fig.\,\ref{fig: CS_err}(a) shows the signal recovery error 
    as a function of the sparsity level $s$ under different levels of hardware variations. 
    %Here we recall that the hardware variation introduces the reading/writing error of $\mathbf C$ in \eqref{eq: sol_x_cs} while mapped to memristor crossbars.  
    We  compare the resulting solution with  the solution obtained from 
   the   orthogonal matching pursuit (OMP) algorithm \cite{tropp2007signal}, a commonly used   software-based CS solver. We observe that the recovery accuracy improves as the signal becomes sparser, namely,     $s$ is smaller. This is not surprising, since a sparser signal can be more stably recovered   at the rate much smaller than
what is commonly prescribed by Shannon-Nyquist theorem \cite{canromtao06}.  By fixing $s$, we  observe that the recovery accuracy decreases while increasing the level of  hardware variations. 
Although the presence of hardware variations negatively affects the recovery accuracy, 
the sparse pattern  error shown by Figs.\,\ref{fig: CS_err}(b) and (c) is acceptable,  as it is  below $6\%$.
% A similar conclusion can be drawn from  Fig.\,\ref{fig: CS_err}(b): the estimation error of sparse pattern increases as $s$ or the level of hardware variation increases.  However, the sparse pattern recovery error is acceptable (below $6\%$).   This is clearly shown by Fig.\,\ref{fig: CS_err}(c), where   we  compare the recovered   signals under $2\%$ and $10\%$ hardware variation with the original   signal and the estimated signal from OMP.
In particular, in Fig.\,\ref{fig: CS_err}(c) the recovered signal yields almost the same sparse support as that of the original signal even in the presence of  $10\%$ hardware variation.
These promising results show that the memristor-based CS solver is quite robust to   hardware variations, and is able to provide  reliable     recovered sparse patterns. 
Lastly, we investigate the convergence of the memristor-based  approach   against different values of the ADMM parameter $\rho$. Similar to Fig.\,\ref{fig: LP_QP_iter}, a moderate choice of $\rho $, namely, $\rho = 10$ in this example, is preferred over  others as shown in Fig.\,\ref{fig: CS_err}(d). 
 
  \section{Power Iteration via Memristors: Application to PCA}
  \label{sec: PCA}
 Principal component analysis (PCA) is the best-known dimensionality-reduction technique to find intrinsic low-dimensional manifolds from high-dimensional data \cite{fodor2002survey}.  The implementation of PCA requires the computation of the principal eigenvalues and the corresponding eigenvectors of a symmetric matrix.  The calculation of eigenvalues and eigenvectors is also motivated by   optimization problems, e.g.,  a  projection onto  semidefinite cones in semidefinite programming \cite{vanboy96}. 
Since power iteration (PI)  is a widely-used algorithm 
for eigenvalue analysis
%that   converges to the eigenvector associated with the largest eigenvalue of the matrix  
\cite{bertsekas1989parallel}, here we describe  a memristor-based PI framework.

 \subsection{Preliminaries on PI}
 PI is an iterative algorithm that      converges to the eigenvector associated with the largest eigenvalue of a matrix. Let $\{ (\lambda_i, \mathbf u_i) \}_{i=1}^n$ denote a set of eigenvalue-eigenvector pairs for matrix $\mathbf A \in \mathbb R^{n \times n}$, where 
 we refer to $\lambda_1$, regardless of its multiplicity, as the dominant eigenvalue.
% without loss of generality  we assume that $\lambda_i \geq \lambda_{i-1}$ for $i=2,3,\ldots, n$. 
 The $k$th iteration of PI is given by
 \cite{golub2012matrix}
 \begin{align}
\mathbf x^{k} = \frac{\mathbf A \mathbf x^{k-1}}{\| \mathbf A \mathbf x^{k-1} \|_2}, \label{eq: power}
\end{align}
where $\mathbf x^0 $ is an arbitrary starting vector. 
If $k \to \infty$, then by \eqref{eq: power}, $\mathbf x^k$ converges 
to the   eigenvector $\mathbf u_1$, and thus 
$
 \frac{(\mathbf x^{k})^T\mathbf A \mathbf x^{k}}{(\mathbf x^{k})^T\mathbf x^{k}}
$ converges to the largest eigenvalue $\lambda_1$. The
convergence   of PI is  geometric, with ratio $\frac{| \lambda_2|}{|\lambda_1|}$ \cite{golub2012matrix}. Therefore, PI
 converges slowly if there is an eigenvalue close in magnitude to the dominant eigenvalue. Moreover, if the largest eigenvalue 
is not unique, say $\lambda_{1} = \lambda_{2}$ with multiplicity $2$,  the limiting point $\mathbf x^{k}$ 
fails to converge to $\mathbf u_1$, and instead 
converges to 
 a linear combination of eigenvectors $\mathbf u_{1}$ and $\mathbf u_{2}$ \cite{panju2011iterative}. 
 Thus, it is required that the  memristor-based PI be able to
%Motivated by that, the desired memristor-based PI method should be able to  
address the issue of repeated eigenvalues.

 \subsection{Memristor-based PI}
 
 It is clear from \eqref{eq: power} that 
  the  PI algorithm involves  a) matrix-vector multiplication $\mathbf A \mathbf x^{k-1}$, and b) evaluation of a vector   norm. Based on \eqref{eq: VIO_ex}, the first operation  is easily implemented using memristor crossbars. And the second operation can be realized using elementary 
  digital (or analog) circuits \cite{liu2017ultra}. 
  The major challenge of customizing PI for  memristor implementation is to determine the multiplicity of the dominant eigenvalue  and to find the corresponding eigenvectors. 
  In what follows, we
    show that with the aid of  Gram-Schmidt process such a problem can be addressed via elementary matrix-vector operations. 
    
    We assume that the largest eigenvalue has multiplicity $s$, namely, $\lambda_{1} = \lambda_{2} = \ldots = \lambda_s$. 
  Under $s$ random initial vectors, we denote by  $\{ \mathbf y_i\}_{i=1}^s$  the  converging vectors of PI. 
  It is known from  \cite{panju2011iterative} that  $\{ \mathbf y_i\}_{i=1}^s$  are linear combinations of eigenvectors $\{ \mathbf u_i\}_{i=1}^s$. This implies two   facts. First, given $p$ initial vectors, the resulting $\{ \mathbf y_i\}_{i=1}^p$ are linearly independent if $p \leq s$ and linearly dependent if  $p > s$. Therefore,  we are able 
   to determine    the number of repeated dominant eigenvalues by 
   adding new columns to $\mathbf Y_p$ until its rank stops increasing
   where
 $\mathbf Y_p \Def [\mathbf y_1, \ldots, \mathbf y_p]$, and its rank can be determined by
 the singularity of $\mathbf Y_p \mathbf Y_p^T$.
  Second, given the number of repeated   eigenvalues, 
 finding the eigenvectors $\{ \mathbf u_i\}_{i=1}^s$ is equivalent to seeking an orthogonal subspace 
   spanned by $\{ \mathbf y_i\}_{i=1}^s$. This procedure is precisely described by the Gram-Schmidt process. 
  Given a sequence of vectors $\{ \mathbf y_i \}_{i=1}^s$,
%(or $\mathbf Y  \Def [\mathbf y_1, \mathbf y_2, \ldots, \mathbf y_s]$),
the Gram-Schmidt process generates a sequence of orthogonal vectors $\{  \mathbf u_i \}_{i=1}^s$ \cite{golub2012matrix}, 
%(or $\mathbf U  \Def [\mathbf u_1, \mathbf u_2, \ldots, \mathbf u_s]$)
\begin{align}
 \mathbf u_i = \mathbf y_i- \sum_{j=1}^{i-1} \frac{\mathbf y_i^T \mathbf u_j}{\mathbf u_j^T \mathbf u_j} \mathbf u_j, ~ i =2,\ldots, s, \label{eq: schmidt}
\end{align}
where $\mathbf u_1 = \mathbf y_1$. 

By incorporating the Gram-Schmidt process \eqref{eq: schmidt}, the generalized  PI algorithm   is able to calculate the dominant eigenvalue even if it is not unique. 
Once the dominant eigenvalue $ \lambda_1 $ is found,   the second largest eigenvalue $\lambda_{2}$ can then be found by performing PI to a new matrix $\mathbf A -\lambda_1  \mathbf u_1 \mathbf u_1^T $, known as a matrix deflation  \cite{golub2012matrix}. 
Since both \eqref{eq: power} and \eqref{eq: schmidt} only involve elementary matrix-vector operations,  it is possible to accelerate PI by using memristors. 

\subsection{Performance evaluation}
\begin{figure}[htb]
\centering 
 \includegraphics[width=0.5\textwidth,height=!]{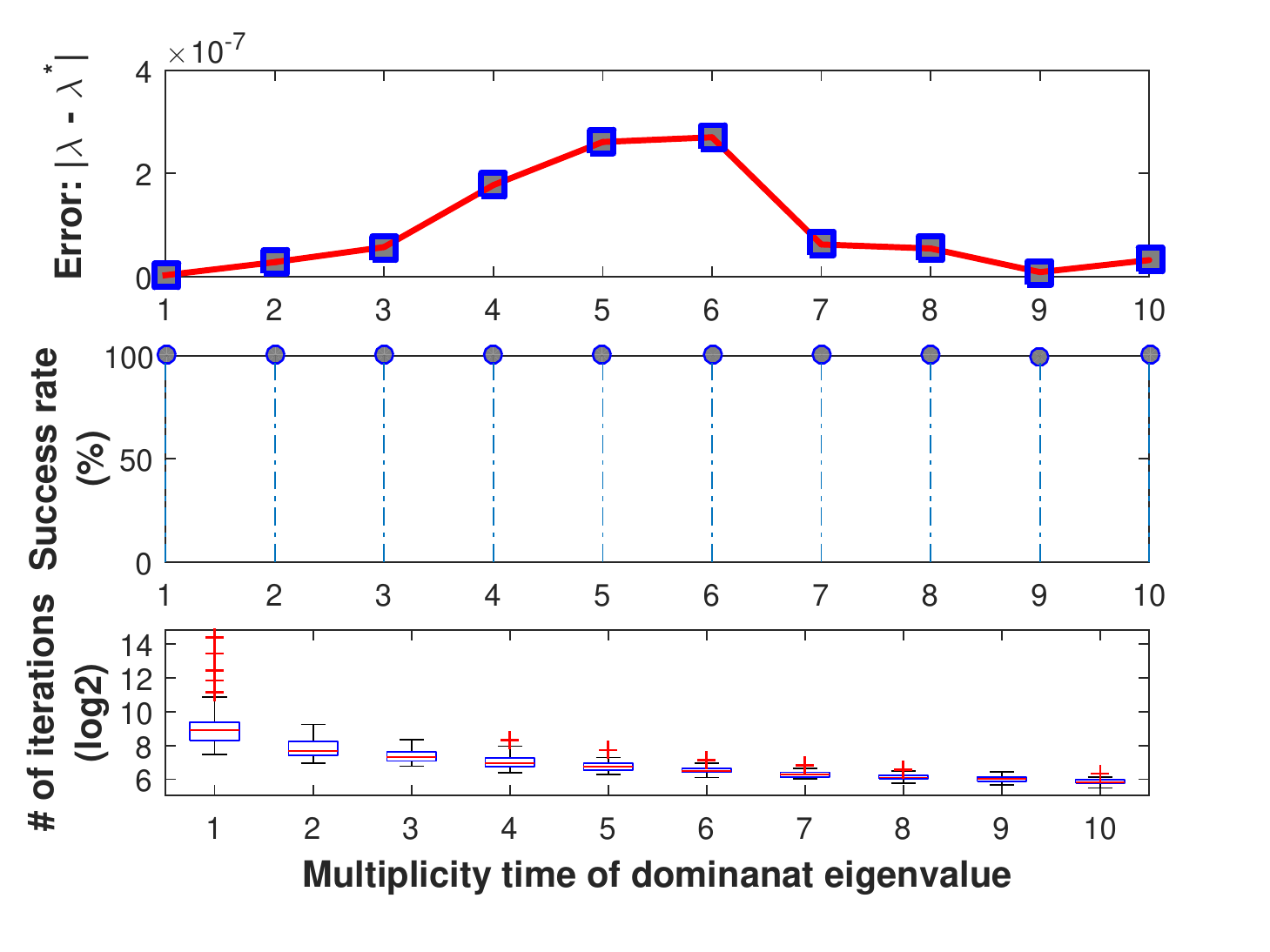}
\caption{Performance of the proposed PI solver against the multiplicity of the dominant eigenvalue.}
\label{fig: PI_err}
\end{figure}

In what follows, we demonstrate the empirical performance of the proposed PI method to compute the dominant eigenvalues/eigenvectors based on a synthetic dataset and to perform PCA based on the   {Iris}   flower dataset   \cite{ben2001support}. 
To specify the eigenvalue problem, let 
 $\mathbf A$ be a symmetric matrix of dimension $n = 50$. We assume that the dominant eigenvalue is repeated $k$ times, where $k \in [1, 10]$. The proposed algorithm continues  until  a  $10^{-4}$-accuracy solution is achieved. Such an experiment is performed over $50$ independent trials. 
 In Fig.\,\ref{fig: PI_err}, we present the computation error, success rate, and the number of iterations of PI against the  multiplicity of the dominant eigenvalue. Here the computation error 
 is averaged over $50$ trials, and  given by the  difference  between the memristor-based  solution $\lambda$ and the optimal solution $\lambda^*$ obtained from the eigenvalue decomposition. As we can see, the proposed PI solver is of high accuracy with error less  than $10^{-6}$. Moreover, at each   trial, the proposed solver correctly    recognizes the number of repeated dominant eigenvalues. And it converges fast, within $1000$ iterations. 
 
\begin{figure}[htb]
  \centering 
 \includegraphics[width=0.5\textwidth,height=!]{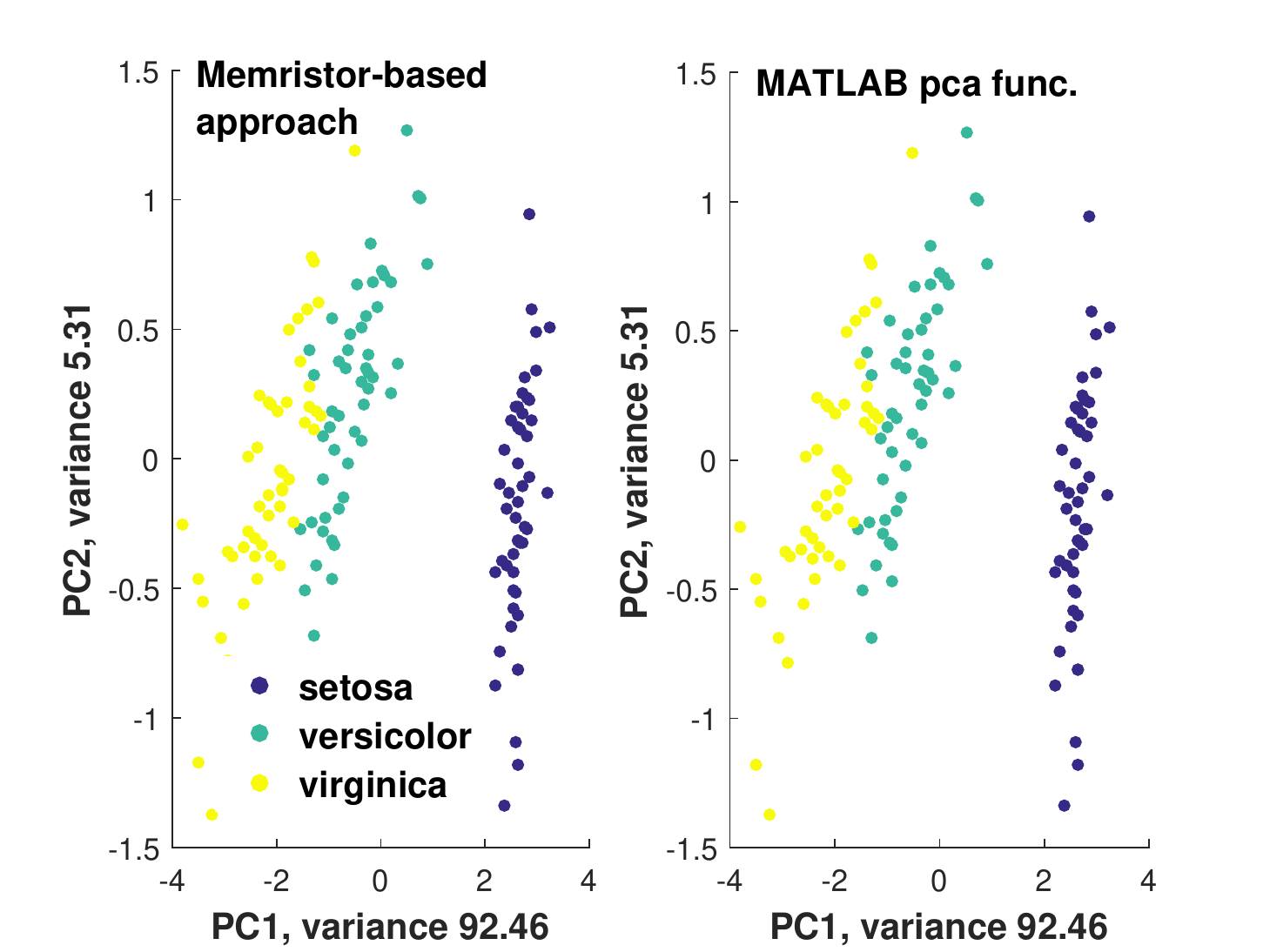}
\caption{PCA  results for the {Iris}   flower dataset. (Left) memristor-based approach; (Right) MATLAB \textit{pca} function.}
\label{fig: PCA}
\end{figure}

 In Fig.\,\ref{fig: PCA}, we apply the proposed PI solver to find the principal components (PCs) of the {Iris}   flower dataset, which contains $150$ iris flowers, and    each flower involves $4$ measurements, sepal length, sepal width, petal length and  petal width. These flowers belong to three different species: setosa,  versicolor, and virginica.
We compare the memristor-based  approach with the standard \textit{pca} function in MATLAB. As we can see, both   methods yield the  same 2D data distribution and the same  variance  of each  PC. These results imply that the application of memristor crossbars is of feasible for this problem. 

\section{Conclusion and Future Directions}
\label{sec: Diss}
In this paper, we 
presented an overview of  a
memristor-based  optimization/computation framework that   exploits both 
memristors' properties and   algorithms' structures.
Popularly used algorithms, ADMM and PI,  
were selected to illustrate  memristor crossbar-based implementations.
%for memristor crossbar-based implementations. 
We showed that ADMM is able to decompose a complex problem  into
matrix-vector multiplications and subproblems for solving systems of linear equations, which then 
facilitates    memristor-based computing architectures. 
To solve the eigenvalue problem using memristor crossbars, 
we   presented a generalized version of  the PI algorithm in the presence of repeated dominant eigenvalues. The effectiveness of memristor-based  framework was illustrated via examples involving LP, QP, compressive sensing and PCA. 
The framework showed a great deal of promise 
  with low computational complexity and high resiliency to hardware variations. %and high scalability for large-scale applications.

%to facilitate  the compact and energy-efficient mapping from algorithm to hardware. 
Although there has been a great deal of progress on the design of memristor-based 
 computation accelerators, many questions and challenges still remain 
 to enable its adoption in  real-life applications, e.g., enhancing  memristor-based computing precision, co-optimizing algorithm and hardware for nonconvex optimization, and
 determining the feasibility of other problems that can benefit from memristor-based hardware implementation. Some specific future directions are discussed below.
 
 % detecting feasibility of problems  in hardware, expanding end-user applications.

First,     memristor-based computing systems have not yet
demonstrated a competitively high computation accuracy for solving practical problems in the presence of hardware variations.
To enhance     precision, extra hardware resources would be needed. 
It is thus essential to optimize a full hardware system under given hardware resources. Problems of interest include selection of device-level components in hardware implementation, and design of energy-efficient on-chip communication infrastructure.

Second,  the convergence of ADMM for nonconvex optimization is not guaranteed. Therefore, new  optimization
algorithms, appropriate for hardware design, are desired to address nonconvex problems, e.g.,   artificial neural network based applications. Traditional algorithms to train neural networks, such as
back-propagation or other gradient-based approaches,  require updating of   the gradient information at each iteration. This leads to frequent writing/reading operations on memristor crossbars and thus an increasing amount of energy consumption.  
Motivated by that, innovation
beyond the existing  algorithms  is encouraged to co-optimize algorithm and hardware for nonconvex optimization. 
 
Third, in many scenarios, it is 
  assumed that certain solutions  exist for the considered optimization and machine learning problems. However, it is possible that the mapped problems on memristor crossbars are infeasible, e.g., no solution exists for an overdetermined linear system. 
   Therefore, a robust memristor crossbar-based solver should be capable of identifying the feasibility of problems. This identification procedure should be  implemented by using device-level components subject to limited   hardware resources.

Fourth, there is much work to be done to expand the applications of memristor crossbars from the end-user perspective. Some  potential lucrative  applications   include   memristor-based smart sensors,   small footprint intelligent
controllers in wearable devices, and on-chip training  platforms in autonomous vehicles and 
Internet of Things. 
% In those use cases, memristor technology provides a solution  to their computing needs, that is,  storing and processing online data with  ultra-fast computation speed. 

To sum up, the memristor technology 
  has the potential to 
  revolutionize computing, optimization and machine learning research due to its orders-of-magnitude improvement in energy efficiency and computation speed.  
  Moving forward, engineers and scientists in different fields, such as, machine learning, signal processing, circuits and systems,  and materials should   
 collaborate with each other to make significant progress on     this exciting research
topic.

%\section{Conclusions}
%\label{sec: conc}
  %%% Try best

\bibliographystyle{IEEEbib}
\bibliography{journal}

% that's all folks
\end{document}